\documentclass{ptephy_v1}

\preprintnumber{XXXX-XXXX}

\usepackage{boites}
\usepackage{times,mathptmx}
\usepackage{amsmath}
\usepackage{bm}
\usepackage{graphicx}
\usepackage{color}

\allowdisplaybreaks[1]

\begin{document}

\title{Theory of antiferromagnetic Heisenberg 
spins on breathing pyrochlore lattice}

\author{\name{\fname{Hirokazu} \surname{Tsunetsugu}}{\ast}}%

\address{%
\affil{1}{The Institute for Solid State Physics, %
The University of Tokyo, Kashiwa 277-8581, Japan}
\email{tsune@issp.u-tokyo.ac.jp}
}

\begin{abstract}%
Spin-singlet orders are studied for the antiferromagnetic Heisenberg 
model with spin $S > \frac{1}{2}$ 
on a breathing pyrochlore lattice, where tetrahedron units 
are weakly coupled and exchange constants have two values $0 < J' \ll J$.  
The ground state has a thermodynamic degeneracy at $J¡Ç=0$,  
and I have studied lattice symmetry breaking associated 
to lifting this degeneracy.   
Third-order perturbation in $J'$ for general spin $S$ shows 
that the effective Hamiltonian has a form of 
three-tetrahedron interactions of pseudospins $\bm{\tau}$,  
which is identical to that previously derived for $S=\frac{1}{2}$, 
and I have calculated their matrix elements for general $S$.  
For this effective Hamiltonian, 
I have obtained its mean-field ground state 
and investigated the possibility of 
lattice symmetry breaking for the cases of $S$=$\frac{3}{2}$ and 1. 
In contrast to the $S$=$\frac{1}{2}$ case, $\bm{\tau}$'s response 
to conjugate field has a $Z_3$ anisotropy in its internal space, 
and this stabilizes the mean-field ground state.  
The mean-field ground state has a characteristic spatial pattern 
of spin correlations related to the lattice symmetry breaking.  
Spin structure factor $S(\mathbf{q})$ is calculated 
and found to have symmetry broken parts with amplitudes of the same 
order as the isotropic part.  
\end{abstract}

\subjectindex{171 Frustration}

\maketitle

\section{Introduction}
\label{sec:intro}

Frustrated magnets are a playground of the experimental and 
theoretical studies for the quest of new quantum phases \cite{FM1,FM2}. 
Thermodynamic degeneracy of the classical ground-state manifold 
is the most important ingredient, 
and the question is how this degeneracy is lifted to select 
a unique quantum ground state if it exists.  
Several cases still select some types orders of magnetic 
dipole moments despite frustration effects, but there also 
exist three other cases from the viewpoint of symmetry breaking: 
(i) Spin rotation symmetry is broken, 
but the order parameter is not a conventional magnetic dipole 
but something more exotic like quadrupole or vector chirality. 
(ii) While spin rotation symmetry is not broken, another 
type of symmetry is broken like lattice symmetry.  
(iii) No symmetry is broken: 
this case corresponds to a spin liquid but 
the liquid behavior does not determine if spin gap is finite or zero.  
Valence bond crystal (VBC) state \cite{Sachdev,Lhuillier} 
is a representative example 
of the case (ii), and the lattice rotation and/or translation 
symmetry is broken. 
Affleck-Kennedy-Lieb-Tasaki (AKLT) state \cite{AKLT} 
is an example of the case (iii) and the spin gap is finite.  

Generally speaking, exotic quantum phases are stabilized 
in frustrated magnets, 
if the frustration is strong enough and also if quantum fluctuations 
are large enough.  
The first condition is related to lattice geometry, and 
the Kagom\'e and pyrochlore lattices are the most typical examples 
in dimensions two and three, respectively.  
As for the second condition, factors enhancing quantum fluctuations 
include a high symmetry of the Hamiltonian,  
a low spatial dimensionality, and a small value of spin $S$.  

The antiferromagnetic Heisenberg model on the pyrochlore lattice 
is a canonical example of frustrated quantum systems 
in three dimensions 
\cite{Reimers,Harris,Canals,Yamashita,Koga,Tsunetsugu1,Tsunetsugu2,Henley}.  
This lattice is a network of corner-sharing tetrahedron units, 
and thus each unit is highly frustrated.  
The pioneering mean-field analysis \cite{Reimers} 
showed a huge degeneracy of the semiclassical ground-state manifold, 
manifested by the presence of zero-energy excitations 
in all over the Brillouin zone \cite{Tsunetsugu3}.  
Over a decade ago I studied the quantum limit $S$=$\frac{1}{2}$ 
of this model and examined the possibility 
of exotic orders like scalar chirality \cite{Tsunetsugu1}.  
This expectation came from the fact that the doubly degenerate 
ground states in each tetrahedron unit have opposite scalar chiralities.  
I derived an effective Hamiltonian in the subspace where 
every tetrahedron unit is within the two-dimensional local ground-state 
multiplet of spin singlet and analyzed that Hamiltonian. 
The result showed that the ground state does not have a scalar 
chirality order but it is a mixture of local singlet dimers 
and/or tetramers and thus the lattice symmetry is broken 
\cite{Tsunetsugu1,Tsunetsugu2}.  
The spatial pattern of these dimers/tetramers is quite complicated 
and this is due to frustration in their configuration.  
Three-quarters of tetrahedron units have a specific 
favorable configuration of dimer pairs, but the remaining quarter 
of units have no favorable configuration.  
This is a frustration in the mean-field level, and quantum fluctuations 
select a uniform order of either dimer pairs or tetramers 
in the remaining part \cite{Tsunetsugu2}.  

In my previous study for the $S$=$\frac{1}{2}$ case, I introduced 
one parameter for controlling geometrical frustration.  
The original pyrochlore lattice was split into two parts:  
one is the set of pointing-up tetrahedron units and the other 
is the set of bonds connecting these units.  
The control parameter is the ratio of exchange constants 
for the two parts $J' /J$, and I approached the original 
model ($J' /J$=1) by perturbation starting 
from the decoupled limit $J' /J$=0 \cite{Tsunetsugu1,Tsunetsugu2}.  
A different split was also examined by another group \cite{Koga} 
but my split had the advantage of keeping the tetrahedral 
lattice symmetry. 

A few years ago, Okamoto et al.~\cite{Okamoto1} noticed that Cr ions in 
the compounds LiGa$_{1-x}$In$_{x}$Cr$_4$O$_8$ ($0 \le x \le 1$) 
constitute a network corresponding to 
my previous perturbative expansion and named this sublattice 
\textit{breathing pyrochlore}.  
Despite the same lattice structure, 
this system differs from my previous model 
in the point that Cr ions have spin $S$=$\frac{3}{2}$.  
The In end ($x=1$) has largest breathing and shows 
an antiferromagnetic phase transition at the temperature $T_N$=13-14 K
\cite{Tanaka,Okamoto2}.  
This is much smaller compared with the modulus of the Weiss temperature 
$|\theta_W |$=332 K, showing the effects of frustration.   
This magnetic phase is very fragile against reducing In-ion concentration, 
disappearing at around $x \sim 0.9$, and the system remains paramagnetic 
down to the lowest temperature 2 K.  
This may indicate that nonmagnetic ground state is stabilized 
in the breathing pyrochlore lattice.
Therefore, it is interesting to reinvestigate the problem now for 
the case of $S$=$\frac{3}{2}$ and also for general $S$, and 
I will examine how the results depend on $S$.  
It will turn out that for $S > \frac{1}{2}$ there appears a generic 
anisotropy in dimer/tetramer configuration and this stabilizes 
a specific spatial modulation of spin correlations.  
This will be manifested in the wave-vector dependence 
of the energy-integrated spin structure factor $S (\mathbf{q})$, 
and I will calculate its explicit form.  

\begin{figure}[tb]
\begin{center}
\includegraphics[width=4.5cm]{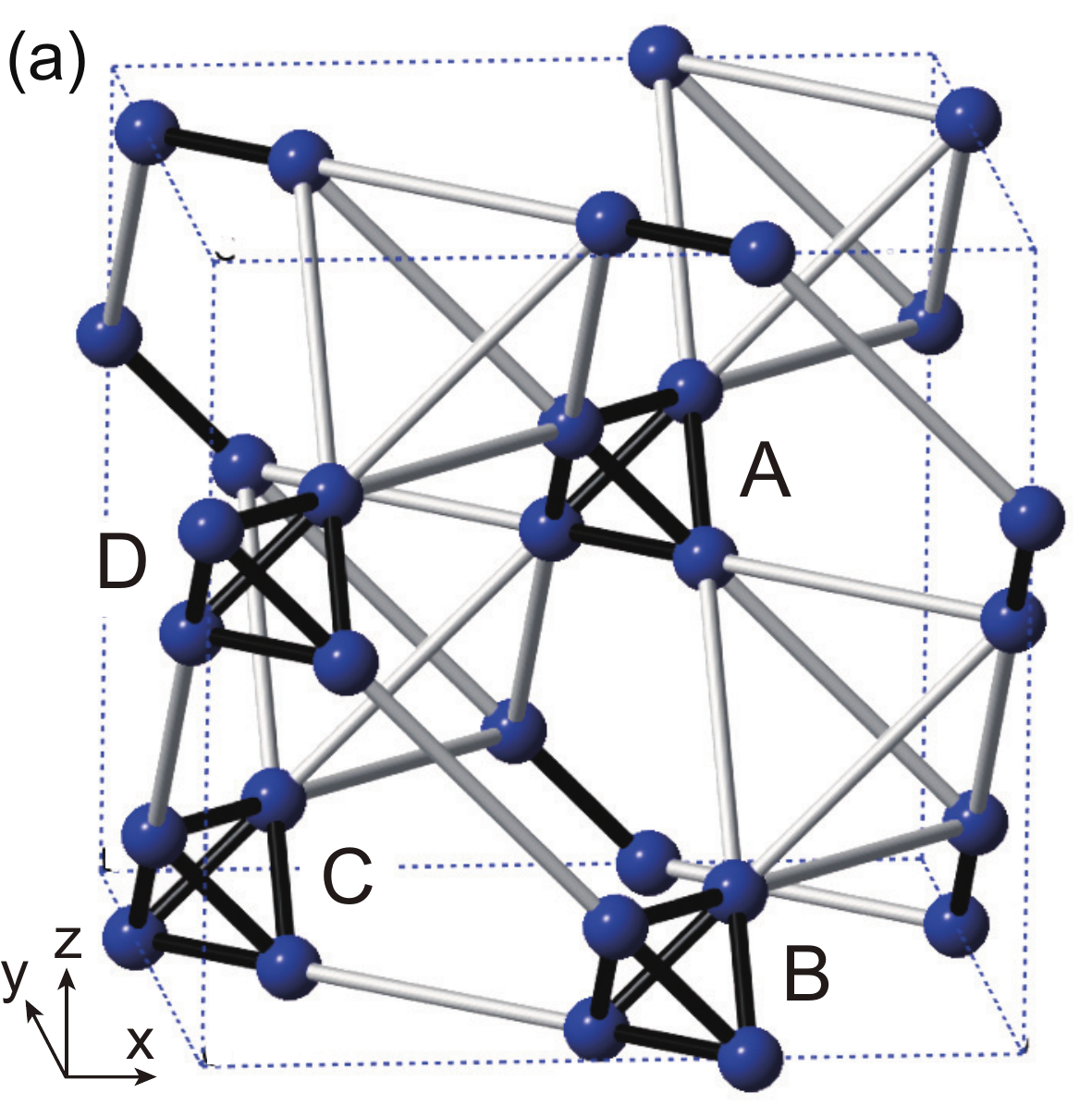}
\hspace{2cm}
\includegraphics[width=4cm]{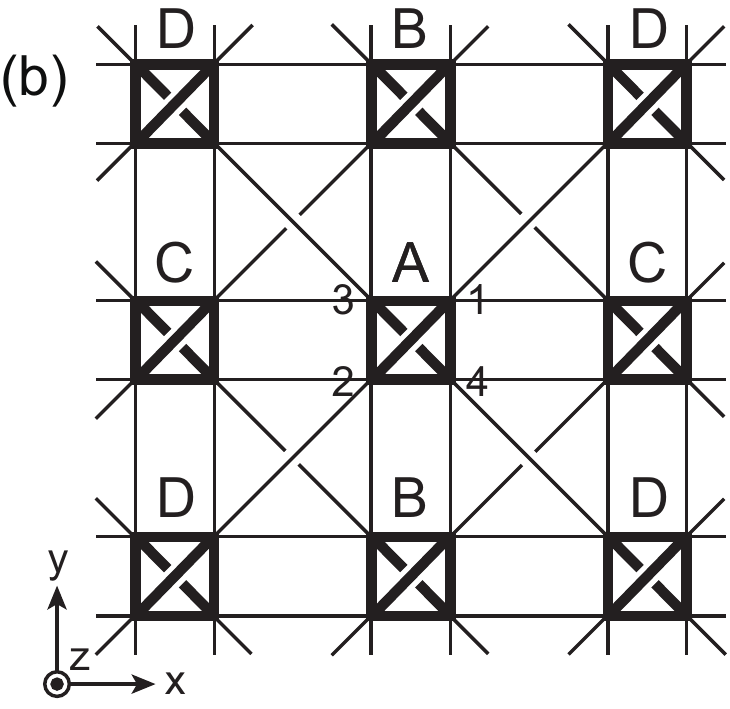}
\end{center}
\caption{
(a) Cubic unit cell of breathing pyrochlore lattice. 
Tetrahedron units made of short bonds (black) with strong $J$ 
are connected by long bonds (gray) with weak $J'$.   
A-D are the sublattice labels of the tetrahedron units.  
(b) A slice of breathing pyrochlore lattice 
projected onto the $xy$-plane.  
The parts shown by thick lines are tetrahedron units. 
The layer of the units A and D is located above that of 
B and C in this part.  
Numbers 1-4 are the site labels in each tetrahedron unit.  
\label{fig:lattice}}
\end{figure}

\section{Model}
\label{sec:model}
In this paper, I study the ground state of a spin-$S$ Heisenberg model 
on a breathing pyrochlore lattice, and apply the results 
to the $S=\frac{3}{2}$ case, which may be realized in the Cr compound 
Li$X$Cr$_4$O$_8$.  
The original pyrochlore lattice is a network of two types 
of corner-sharing tetrahedra, and they have the same size.  
In breathing pyrochlore lattice, 
one type of tetrahedra expand in size and the 
other type shrink, but neither of them change 
their shape of regular tetrahedron.  
See Fig.~\ref{fig:lattice}(a).  
The Hamiltonian to study is a spin-$S$ Heisenberg model with 
antiferromagnetic interactions between nearest neighbor sites 
on the breathing pyrochlore lattice, and I will analyze the cases 
of $S=\frac{3}{2}$ and 1 in detail.  
Exchange coupling has two values depending 
on bond length in the breathing lattice structure 
\begin{subequations}
\begin{align}
H &= J \sum_{\mathbf{r}} \sum_{1 \le i < j \le 4} \!\!
\mathbf{S}_i (\mathbf{r}) \cdot \mathbf{S}_j (\mathbf{r}) 
+ J' \hspace{-0.4cm} \sum_{\langle (i, \mathbf{r}), (j,\mathbf{r}') \rangle} 
\!\!\!\! \mathbf{S}_i (\mathbf{r}) \cdot \mathbf{S}_j (\mathbf{r}') 
\label{eq2:modelA}
\\
&= \sum_{\mathbf{r}} H_{\mathrm{unit}} (\mathbf{r}) 
+ \sum_{\langle \mathbf{r}, \mathbf{r}' \rangle} 
H_{1} (\mathbf{r}, \mathbf{r}')  , 
\label{eq2:model}
\end{align}
\end{subequations}
where $\mathbf{r}$ denotes the position of small tetrahedron 
and $i,j$ are the site labels as shown in Fig.~\ref{fig:lattice}(b).  
Six bonds in each small tetrahedron 
unit have strong interaction $J > 0$, while the units are 
coupled by long bonds with weak antiferromagnetic interaction 
$ 0 < J' (\ll J)$.  
In the following, I will call small tetrahedra with 
strong bonds \textit{tetrahedron units}, and then 
they are connected by weak bonds in large tetrahedra.  
Throughout this paper, I use $N$ to denote the number of 
spins, and then the number of tetrahedron units 
is $\frac{1}{4} N$. 
Note that the Hamiltonian is invariant 
upon exchanging $J$ and $J'$ due to 
the lattice symmetry.  

I studied in Refs.~\cite{Tsunetsugu1,Tsunetsugu2}
this model for the $S=\frac{1}{2}$ case 
and discovered a complex pattern of spin dimers and tetramers  
in the ground state.  
I will later focus on the special cases of $S=\frac{3}{2}$ and $S=1$ 
afterwards, but first 
study this Hamiltonian for general $S$  
to compare the cases of different spin quantum numbers $S$. 

\section{Spin singlet states in one tetrahedron unit}
\label{sec:singlets}
Following Refs.~\cite{Tsunetsugu1,Tsunetsugu2}, 
I employ an approach of degenerate perturbation in $J'/J$. 
The first step is to understand eigenstates 
in the limit of decoupled tetrahedra ($J'$=0).  
The Hamiltonian of a single tetrahedron unit at position $\mathbf{r}$ is 
\begin{equation}
H_{\mathrm{unit}} (\mathbf{r}) = J \! \sum_{1 \le i < j \le 4} 
\mathbf{S}_i (\mathbf{r}) \cdot \mathbf{S}_j (\mathbf{r}) 
= J [ {\textstyle \frac{1}{2}} \, \mathbf{S}_{\mathrm{unit}}^2 (\mathbf{r}) 
- 2 S (S+1) ]  . 
\label{eq2:unit}
\end{equation} 
Here, $\mathbf{S}_{\mathrm{unit}} (\mathbf{r}) 
= \sum_{i=1}^4 \mathbf{S}_i (\mathbf{r}) $ is 
the total spin of this unit and its quantum number is an 
integer $0 \le S_{\mathrm{unit}} \le 4S $.  
Therefore, the eigenenergies of $H_{\mathrm{unit}} (\mathbf{r})$ 
are determined by the unit spin alone as 
$E_{\mathrm{unit}} = 
J [ {\textstyle \frac{1}{2}} \, 
S_{\mathrm{unit}} (S_{\mathrm{unit}} +1) - 2 S (S+1) ]$ 
and the ground-state manifold 
coincides with the entire space of $S_{\mathrm{unit}}$=0.  
These results have been well-known including the fact that 
the ground states in each unit have degeneracy $2S+1$, 
and thus the ground-state degeneracy in the entire system is 
$(2S+1)^{N/4}$ where $N$ is the number of original spins \cite{Tsunetsugu1}.   
This corresponds to the residual entropy that is precisely 
25\% of the total entropy irrespective of the value of $S$.  

The issue of this study is how the weak inter-unit interactions $J'$ 
release the macroscopic entropy and which type of ground state is selected.  
Since anything interesting happens in the spin-singlet space, 
the spin rotation symmetry has no chance to be broken, 
and it is the lattice symmetry that can be broken.  
To examine this issue, symmetry argument is useful and 
I will check how the $(2S+1)$-fold ground states in each unit 
are transformed with operations of the point group symmetry.   
Each unit has the tetrahedral symmetry $T_d$, 
and this point group has 5 types of irreducible representations (irreps) 
\cite{group}:
2 one-dimensional ones (A$_1$ and A$_2$), 
1 two-dimensional one (E), and 
2 three-dimensional ones (T$_1$ and T$_2$).  

To perform calculation, we need an explicit form of 
$(2S+1)$-states in the space of $S_{\mathrm{unit}}=0$.
A convenient choice is the following one 
\begin{equation}
\Phi_{l} = 
\sum_{l_z=-l}^{l} 
C(l,l_z) 
\phi^{(12)} (l,l_z ) \otimes 
\phi^{(34)} (l,-l_z ) , \ \ 
\mathbf{S}_{\mathrm{unit}}^2 \Phi_{l} = 0 , 
\label{eq2:CG4}
\end{equation}
where $l$ is an integer satisfying $0\,$$\le$$\,l$$\,\le$$\, 2S$.  
Here, $\phi^{(12)} (l,l_z )$ is the wavefunction 
in which the two spins $\mathbf{S}_1$ and $\mathbf{S}_2$ 
couple to form a state with the composite spin $l$ and 
its $z$-projection $l_z$, while 
the remaining two spins $\mathbf{S}_3$ and $\mathbf{S}_4$ do the same 
in $\phi^{(34)} (l,-l_z )$ but with the opposite $z$-projection.  
These two-spin wavefunctions are written 
with single-spin bases as 
\begin{equation}
\phi^{(ij)} (l,l_z ) = \sum_{m=\max \{-S, l_z - S \} }^{\min \{S, l_z + S \} }
\langle S,S,m,l_z-m | l,l_z \rangle 
|S,m \rangle_i \otimes |S,l_z -m \rangle_j , 
\label{eq2:2spinwf}
\end{equation}
where $\langle S,S,m,l_z-m | l,l_z \rangle $ is the Clebsch-Gordan (CG) 
coefficient\footnote[1]{
Here, the Clebsch-Gordan coefficient is defined as 
$\langle j_1 , j_2 , m_1 , m_2 | J M \rangle$ for 
the combination of two angular momenta, 
$\mathbf{j}_1 + \mathbf{j}_2 = \mathbf{J}$.  
$m_{1,2}$ and $M$ are the $z$-component of 
$\mathbf{j}_{1,2}$ and $\mathbf{J}$, respectively. 
}
of combining two angular momenta \cite{Messiah}, and these wavefunctions 
have the symmetry 
\begin{equation}
\phi^{(ji)} (l,l_z ) = (-1)^{2S+l} \phi^{(ij)} (l,l_z ).   
\label{eq:sym1}
\end{equation}
These two composite spins couple and finally form a total spin singlet 
in a tetrahedron unit.  
The prefactor $C$ in Eq.~(\ref{eq2:CG4}) is also given by a CG coefficient 
and this is simple because the total spin is singlet 
\begin{equation}
C(l,l_z ) = \langle l,l,l_z,-l_z | 0,0 \rangle 
= \frac{(-1)^{l-l_z}}{\sqrt{2l+1}} = C(l,-l_z ) . 
\label{eq:CG}
\end{equation}
These $(2S+1)$ wavefunctions $\{ \Phi_{l} \}$ constitute 
a complete orthonormal set in the $S_{\mathrm{unit}}$=0 space at each 
tetrahedron unit.  

I have classified these $\Phi_l$'s according to the 
$T_d$ point group symmetry.  
This was done by calculating the characters of the symmetry 
operations, and 
the details of calculation are explained in Appendix \ref{sec:A1}.  
The result turns out interesting. 
For any value of individual spin $S$, all of the $2S+1$ ground states 
in the tetrahedron unit belong to $A_1$, $A_2$, and $E$ 
irreps,  
while no ground states transform as the three-dimensional irrep 
$T_{1}$ or $T_{2}$.  
I have calculated the multiplicity of these irreps 
and the result is 
$\bigl\{ \Phi_{l} \bigr\}_{l =0}^{2S} 
= n_{1 +} A_1 \oplus n_{1-} A_2 \oplus n_2 E$ 
with 
\begin{equation}
n_{1 \pm}= {\textstyle \frac16} ( 2S+1 \pm 3\chi_2 + 2\chi_5 ) , \ 
n_2 = {\textstyle \frac{1}{3}} ( 2S+1 - \chi_5 ) .
\label{eq2:multiplicity}
\end{equation}
Here, $\chi_2 $=$\mbox{mod } (2S-1 , 2)$ and 
$\chi_5$=$\mbox{mod } (2S-1 , 3)-1$. 

\section{Effective Hamiltonian for general $S$}
\label{sec:effective}
Now, I am going to derive an effective Hamiltonian that 
lifts the macroscopic degeneracy of the ground states 
in the limit of decoupled tetrahedra.  
To this end, one needs a degenerate perturbation in the 
weak interaction $J'$, and I succeeded in this task 
for the $S=\frac{1}{2}$ case after a lengthy calculation 
of many matrix elements \cite{Tsunetsugu1,Tsunetsugu2}. 
For larger spins, 
the local Hilbert space increases its dimension, and 
this makes calculations more impracticable and difficult. 
Therefore, I took a different strategy and tried to 
simplify the formulation in perturbation as much as possible. 
I have achieved a huge simplification in the third-order 
perturbation, and this works for any value of spin $S$.  
With this simplified formulation, my previous result for the $S=\frac{1}{2}$ 
case is also easily reproduced.  

The effective Hamiltonian is to be derived for describing 
dynamics in the low-energy subspace where all the local 
states at tetrahedron units are within the spin-singlet manifold, 
and let me comment on its validity.  
The use of such an effective Hamiltonian is justified 
under two conditions.  
The first condition is the presence of a finite spin gap $\Delta_s > 0$. 
The size of the spin gap depends on the ratio of two 
exchange constants\footnote[2]{%
The spin gap is $\Delta_s = J$ in the decoupling limit $J'=0$.  
At small $J'/J$, the ground-state energy has a correction 
starting from the order $J'^2/J$, while the $S_{\mathrm{unit}}=1$ state 
can hop from one tetrahedron to neighboring ones 
with matrix element proportional to $J'$. 
Therefore, the leading correction in the spin gap 
is the order $J'$, and 
$\Delta_s = J - a(S) J' + \cdots$.  
}, 
$\Delta_s = J \bar{\Delta}(\frac{J'}{J};S)$ 
with $\bar{\Delta}(0;S)=1$, 
and the first condition is satisfied at least 
for small $\frac{J'}{J}$.  
The second condition is that the energy range of consideration 
should be smaller than the spin gap, 
$\varDelta E < \Delta_s$, where 
$\varDelta E$ is measured from the ground-state energy.  
For studying the high-energy region $\varDelta E > \Delta_s$, 
it is necessary to take account of the subspaces with 
$S_{\mathrm{unit}} \ge 1$ on the same footing as 
the $S_{\mathrm{unit}} =0$ subspace, 
which is beyond the approximation of this effective Hamiltonian. 

Before demonstrating how perturbation calculation is simplified, 
I now introduce matrix elements necessary for that 
and discuss their symmetry.  
They are two-spin correlations defined for a pair of ground states 
in one tetrahedron unit\footnote[3]{%
The result that the matrix elements are 
diagonal in spin space does not depend on the choice of 
basis states.  
For example, one can use those in Eq.~(\ref{eq2:CG4}) 
or bases of the irreps of $T_d$ group.
}
\begin{equation}
\langle \Phi_\alpha | S_i^\mu S_j^\nu | \Phi_\beta \rangle
= {\textstyle \frac{1}{3}} \delta_{\mu \nu} f_{\alpha \beta}^{(ij)}, \ \ 
f_{\alpha \beta}^{(ij)}=\langle \Phi_\alpha | 
\mathbf{S}_i \cdot \mathbf{S}_j | \Phi_\beta \rangle , 
\label{eq:Fmat}
\end{equation}
Here, $\mu, \nu \in \{ x, y, z \}$ are spin index, and 
the result is diagonal in spin space because 
spin-singlet states are rotationally invariant. 
For the same site correlation, it is trivial, 
simply $f_{\alpha \beta}^{(ii)}$=$S(S+1) 
\langle \Phi_\alpha | \Phi_\beta \rangle$.   
Using the fact that the two states are both spin singlet, 
one can further prove important symmetries of $f^{(ij)}$ for $i$$\ne$$j$.
For a site pair in a tetrahedron unit, let call the remaining 
two sites its \textit{conjugate site pair}; 
\textit{e.g.}, for the site pair 1-2, its conjugate pair is 3-4.  
For the site pair $i$-$j$, let us define 
\begin{equation}
f_{\alpha \beta}^{(ij)} =: - c_0 
\langle \Phi_\alpha | \Phi_\beta \rangle 
+ \bigl( {\mathsf{F}}^{(ij)} \bigr)_{\alpha \beta},  \ \ 
c_0 = {\textstyle \frac{1}{3}} S(S+1) , 
\label{eq:Fdef}
\end{equation}
and then $\mathsf{F}^{(ij)}$ is identical to the value 
for its conjugate site pair: 
\begin{equation}
{\mathsf{F}}^{(12)}={\mathsf{F}}^{(34)}, \ 
{\mathsf{F}}^{(13)}={\mathsf{F}}^{(24)}, \ 
{\mathsf{F}}^{(14)}={\mathsf{F}}^{(23)}, 
\label{eq:PairEquiv}
\end{equation}
and the sum of these vanish
\begin{equation}
{\mathsf{F}}^{(12)}+{\mathsf{F}}^{(13)}+{\mathsf{F}}^{(14)}=0. 
\label{eq:Neutral}
\end{equation}
These $\mathsf{F}$'s are a square matrix with dimension $2S+1$.  
These relations (\ref{eq:PairEquiv}) and (\ref{eq:Neutral}) 
will be referred to in the following 
as {\em conjugate-pair equivalence} and {\em neutrality identity}, 
respectively.  

Before going to perturbation calculations, 
I quickly prove Eqs.~(\ref{eq:PairEquiv}) and (\ref{eq:Neutral}). 
For the conjugate-pair equivalence, it is sufficient to 
prove $f_{\alpha \beta}^{(12)}$=$f_{\alpha \beta}^{(34)}$, 
and the following proof does not depend on 
the choice of site pair $i$-$j$.  
Let $\mathbf{S}_{\mathrm{unit}}$ be the total spin 
in the tetrahedron unit, 
$\mathbf{S}_{\mathrm{unit}}$=$\sum_{i=1}^{4} \mathbf{S}_i$, 
and here I drop the label of the unit position $\mathbf{r}$, since 
all the calculations are limited in one unit. 
For any tetrahedron singlet state $\Phi_{\alpha}$, 
the most important relation is 
\begin{equation} 
\mathbf{S}_{\mathrm{unit}} | \Phi_\alpha \rangle = \mathbf{0} , \ \ 
\langle \Phi_\alpha | \mathbf{S}_{\mathrm{unit}} = \mathbf{0} . 
\label{eqB:S0}
\end{equation} 
Another relation to use is the identity 
$\mathbf{S}_i \cdot \mathbf{S}_j = 
\frac{1}{2} \mathbf{S}_{ij}^2 - S(S+1)$, 
where $\mathbf{S}_{ij} \equiv \mathbf{S}_{i} + \mathbf{S}_{j}$ 
is the composite spin of the pair.  
Then, the relation to prove is equivalent to 
\begin{equation}
\langle \Phi_\alpha | \mathbf{S}_{12}^2 | \Phi_\beta \rangle 
= \langle \Phi_\alpha | \mathbf{S}_{34}^2 | \Phi_\beta \rangle . 
\label{eqB:f12f34}
\end{equation}
Using $\mathbf{S}_{\mathrm{unit}}$, the 3-4 site pair can be represented by 
quantities related to the 1-2 site pair
\begin{equation}
\langle \Phi_\alpha | \mathbf{S}_{34}^2 | \Phi_\beta \rangle 
= \langle \Phi_\alpha | (\mathbf{S}_\mathrm{unit} - \mathbf{S}_{12} )
\cdot (\mathbf{S}_\mathrm{unit} - \mathbf{S}_{12} ) | \Phi_\beta \rangle 
= \langle \Phi_\alpha | \mathbf{S}_{12}^2 | \Phi_\beta \rangle . 
\label{eqB:pairs}
\end{equation}
This completes the proof of the conjugate-pair equivalence.  

Using the conjugate-pair equivalence for three pairs, one can rewrite 
the neutrality identity as follows 
\begin{equation} 
\Bigl\langle \Phi_\alpha \Big| S(S+1) + {\textstyle \frac{1}{2}} 
\sum_{1 \le i < j \le 4} \mathbf{S}_{i} \cdot \mathbf{S}_{j} 
\Big| \Phi_\beta \Bigr\rangle = 0.  
\end{equation} 
It is straightforward to prove this, 
since the left-hand side is nothing but 
$\langle \Phi_\alpha | \frac{1}{4} 
\mathbf{S}_\mathrm{unit}^2 
| \Phi_\beta \rangle$=0.   
Thus, the neutrality identity is also proved.  

\begin{figure}[tb]
\begin{center}
\includegraphics[height=3.5cm]{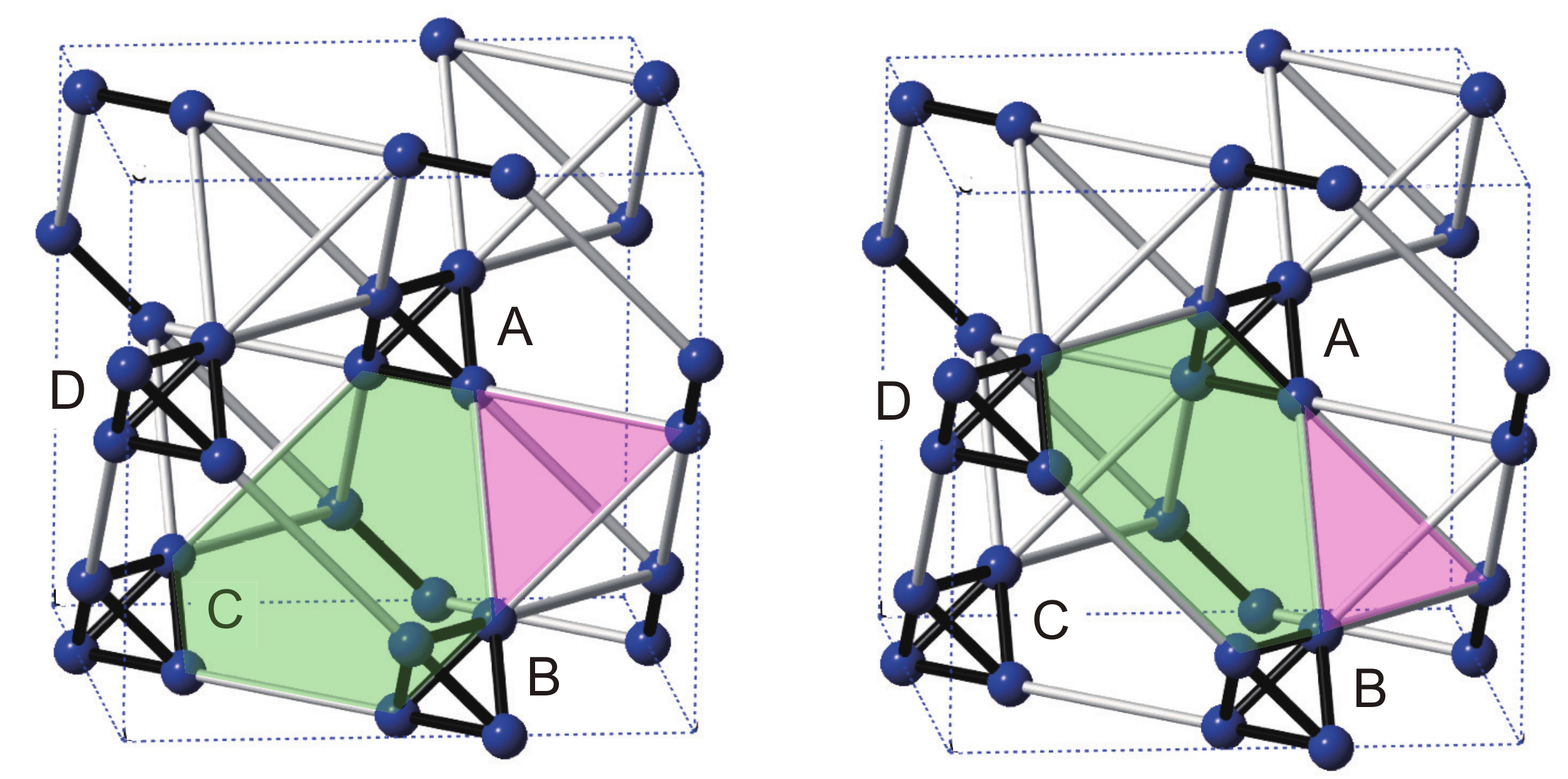}
\end{center}
\caption{Two types of paths appearing in the third-order perturbation. 
Shown are 2 pairs of hexagon and triangular loops that 
share the same long bond connecting A and B units. 
Each loop contains 3 long bonds.  
}
\label{fig:loops}
\end{figure}

\begin{figure}[tb]
\begin{center}
\includegraphics[width=13cm]{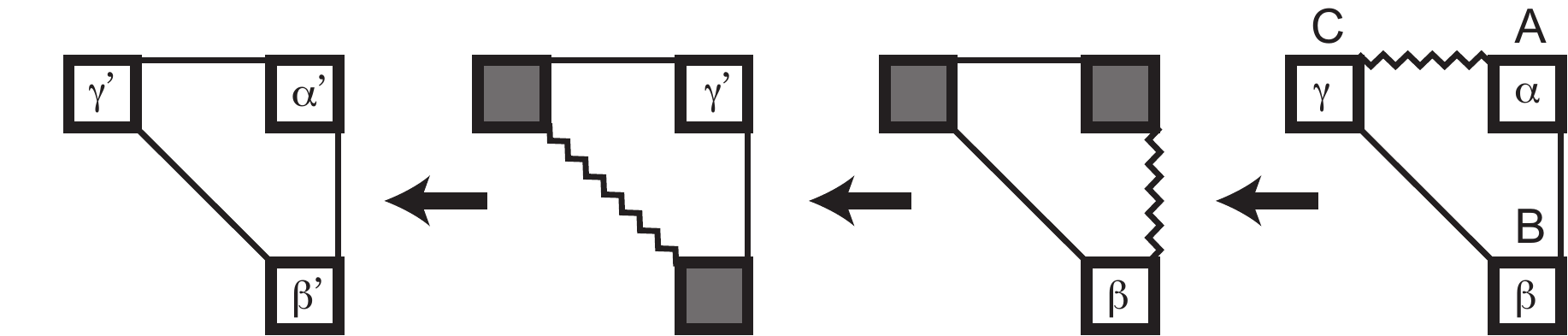}
\end{center}
\caption{
One process in the third-order perturbation for 
the tetrahedron units $A$-$C$. 
This starts from the product state of three singlets  
$\big| \Phi_{\alpha}^{A} \Phi_{\beta}^{B} \Phi_{\gamma}^{C} \bigr\rangle$, 
and each zigzag bond depicts a perturbation 
$J' S_i (\mathbf{r}) \cdot S_j (\mathbf{r}')$ to be operated.  
Shadowed units are tetrahedra excited to $S_{\mathrm{unit}}$=1, 
while all the others remain in $S_{\mathrm{unit}}$=0.  
Different orders of operating three perturbations generate 
5 other processes, and all of them have 
the same contribution as the process shown here, since 
the three perturbations commute to each other.  
\label{fig:pert3}}
\end{figure}

We are now ready to start a degenerate perturbation for 
constructing an effective Hamiltonian.  
In perturbation in $J'$, first-order terms vanish and 
second-order terms only yield a constant energy shift for all states
\begin{equation}
\varDelta E_{0}^{(2)} 
= - \frac{J'^2}{6J} \bigl[ S(S+1) \bigr]^2 \times \frac{3}{2} N , 
\label{eq2:E2nd}
\end{equation} 
where $\frac{3}{2} N$ is the total number of long bonds. 
Therefore, the leading terms that lift the degeneracy 
are third order ones, which was explicitly derived 
for the $S=\frac{1}{2}$ case~\cite{Tsunetsugu1}.  
Beware that there are two types of third-order terms. 
One corresponds to perturbation paths on triangular loops 
made of weak bonds alone, 
and they contribute only a constant energy shift again, 
\begin{equation}
\varDelta E_{0}^{(3)} 
= \frac{J'^3}{6J^2} \bigl[ S(S+1) \bigr]^3 \times N , 
\label{eq2:E3rd}
\end{equation} 
where $N$ should be read as the number of the triangular loops, 
which is identical to the number of original spins.

The other type is what we need and corresponds to perturbation 
paths on hexagon loops each of which includes three weak bonds. 
Two examples of the latter type are shown in Fig.~\ref{fig:loops}. 
The colored hexagon loop in the left panel includes 
the three tetrahedron units $ABC$ 
and the corresponding third-order perturbation term is given by 
\begin{subequations}
\begin{align}
&\bigl\langle \Phi_{\alpha '}^A \Phi_{\beta '}^B \Phi_{\gamma '}^C \big| 
H_{\mathrm{eff}}(ABC) 
\big| \Phi_{\alpha}^{A} \Phi_{\beta}^{B} \Phi_{\gamma}^{C} \bigr\rangle 
\nonumber\\
&= {3! J'^3} \sum_{v_1, v_2} 
\frac{\bigl\langle \Phi_{\alpha '}^A \Phi_{\beta '}^B \Phi_{\gamma '}^C \big| 
 \mathbf{S}_4 (A) \cdot \mathbf{S}_1 (B) 
\big| v_2 \bigr\rangle \bigl\langle v_2 \big| 
 \mathbf{S}_3 (B) \cdot \mathbf{S}_4 (C) 
\big| v_1 \bigr\rangle }
{\varDelta \varepsilon_{ABC} (v_2 ) \, \varDelta \varepsilon_{ABC} (v_1) }
\nonumber\\
&\hspace{4cm} \times \bigl\langle v_1 \big| 
 \mathbf{S}_1 (C) \cdot \mathbf{S}_3 (A) 
\big| \Phi_{\alpha}^{A} \Phi_{\beta}^{B} \Phi_{\gamma}^{C} \bigr\rangle 
\label{eq:pert3a}
\\
&=\frac{3! J'^3}{(2J)^2} 
\Bigl\langle \Phi_{\alpha '}^A \Phi_{\beta '}^B 
\Phi_{\gamma '}^C \Big| 
[ \mathbf{S}_4 (A) \cdot \mathbf{S}_1 (B) ]
[ \mathbf{S}_3 (B) \cdot \mathbf{S}_4 (C) ]
[ \mathbf{S}_1 (C) \cdot \mathbf{S}_3 (A) ]
\Big| \Phi_{\alpha}^{A} \Phi_{\beta}^{B} \Phi_{\gamma}^{C} \Bigr\rangle
\label{eq:pert3b}
\\
&=\frac{3! J'^3}{(2J)^2} \sum_{\mu_1,\mu_2,\mu_3} 
\Bigl[ \bigl\langle \Phi_{\alpha '} \big| 
 {S}_4^{\mu_1} 
 {S}_3^{\mu_3} 
\big| \Phi_{\alpha} \bigr\rangle \Bigr]_A 
\Bigl[ \bigl\langle \Phi_{\beta '} \big| 
 {S}_1^{\mu_1} 
 {S}_3^{\mu_2} 
\big| \Phi_{\beta} \bigr\rangle \Bigr]_B
\Bigl[ \bigl\langle \Phi_{\gamma '} \big| 
 {S}_4^{\mu_2} 
 {S}_1^{\mu_3} 
\big| \Phi_{\gamma} \bigr\rangle \Bigr]_C
\label{eq:pert3c}
\\
&= \frac{J'^3}{6J^2} \, 
{f}^{(12)}_{A,\alpha' \alpha} \, 
{f}^{(13)}_{B,\beta'  \beta} \, 
{f}^{(14)}_{C,\gamma' \gamma} , 
\label{eq:pert3d}
\end{align}
\end{subequations}
where $v$'s are excited states of the system of the units $ABC$ 
and $\varDelta \varepsilon_{ABC} (v)$ is their excitation 
energy measured from the ground state value.  
Note that the conjugate-pair equivalence has been used for the unit $A$, 
$f^{(34)}=f^{(12)}$.  
One process is depicted in Fig.~\ref{fig:pert3} and 
this corresponds to the matrix elements in Eq.~(\ref{eq:pert3a}).  
The factor $3!$ comes from the fact that 
different orders of three $\mathbf{S}\cdot\mathbf{S}$'s 
have all the same contribution.  
The sums $\sum_{v_1, v_2}$ are originally taken over all the excited states 
of the units $ABC$,  
but matrix elements are nonvanishing 
only with those $v$'s in which two units 
are excited to $S_{\mathrm{unit}}$=1 
and one unit remains $S_{\mathrm{unit}}$=0.  
Therefore, the excitation energy is always 
$\varDelta \epsilon_{ABC} (v)$=$2J$ for any of those $v$, 
and one can move the energy denominators to outside the sum.  
I should emphasize that this is a special feature of the present model, 
and this simplifies calculations.  
After doing this, one can modify the sum $\sum_{v_1, v_2}$ 
now with including the ground states.  
This modification does not change the result, 
because the additionally included ground states have only 
zero matrix elements.\footnote[4]{
Recall $\langle \Phi_{\beta} | S_j^\mu | \Phi_{\alpha} \rangle$=0 
for any pair of states in the $S_{\mathrm{unit}}$=0 subspace.  
This is because $S_j^\mu | \Phi_{\alpha} \rangle$ belongs 
to the $S_{\mathrm{unit}}$=1 subspace.
}
The modified sum for the virtual states is taken 
over the entire Hilbert space 
of the three units, and therefore 
we can safely drop this sum, 
$(\sum_{\mathrm{excited\ states}} + \sum_{\mathrm{ground\ states}} )
|v \rangle \langle v| = 1$. 
Thus, Eq.~(\ref{eq:pert3b}) is obtained, and 
it is straightforward to rewrite that into 
the final step (\ref{eq:pert3d}).  
There, the unit index $A$-$C$ is added in the subscript 
to show explicitly which unit contributes to each $f$ factor.

\begin{figure}[bt]
\begin{center}
\includegraphics[width=5.0cm]{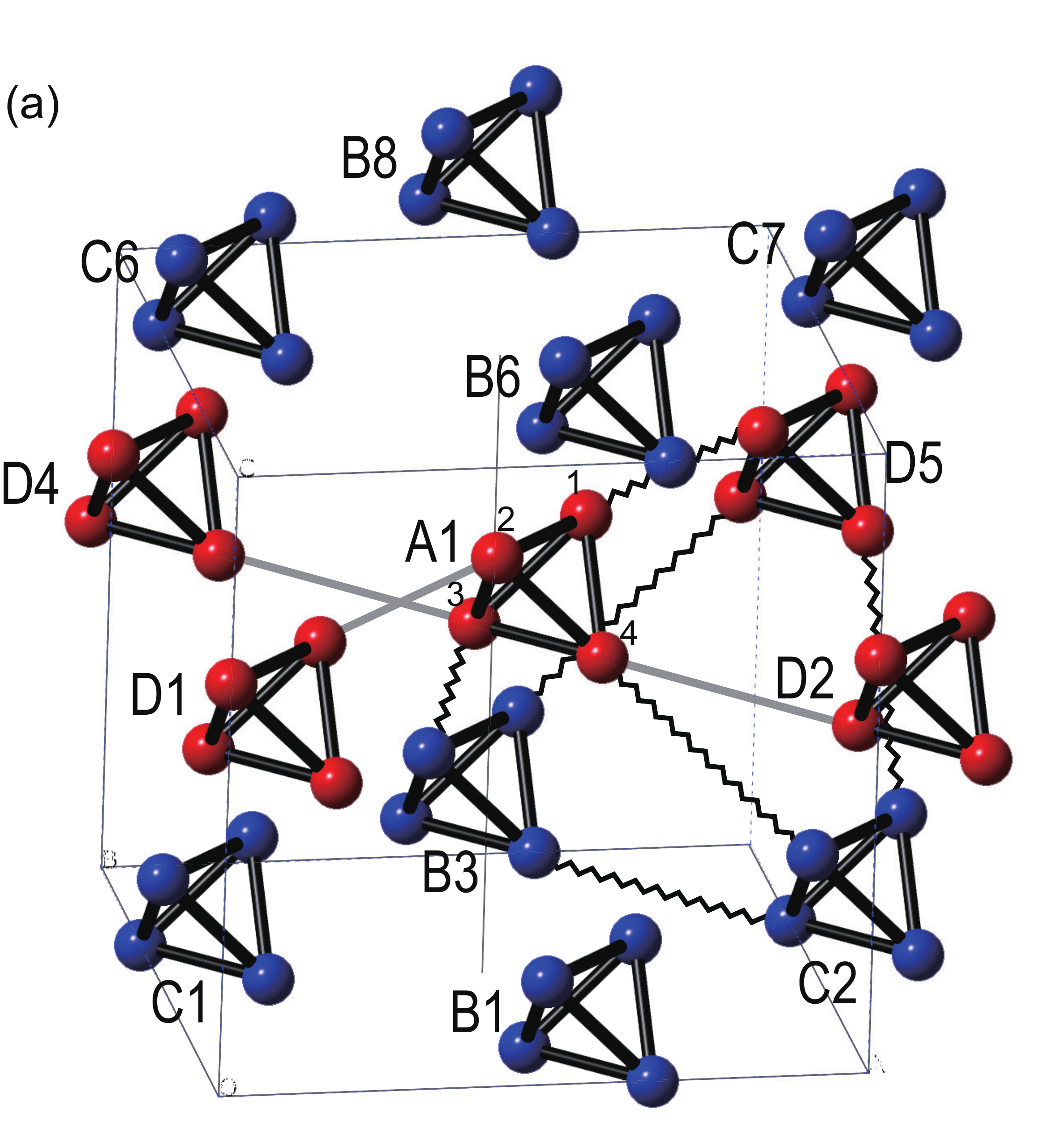}
\hspace{1.5cm}
\includegraphics[width=4cm]{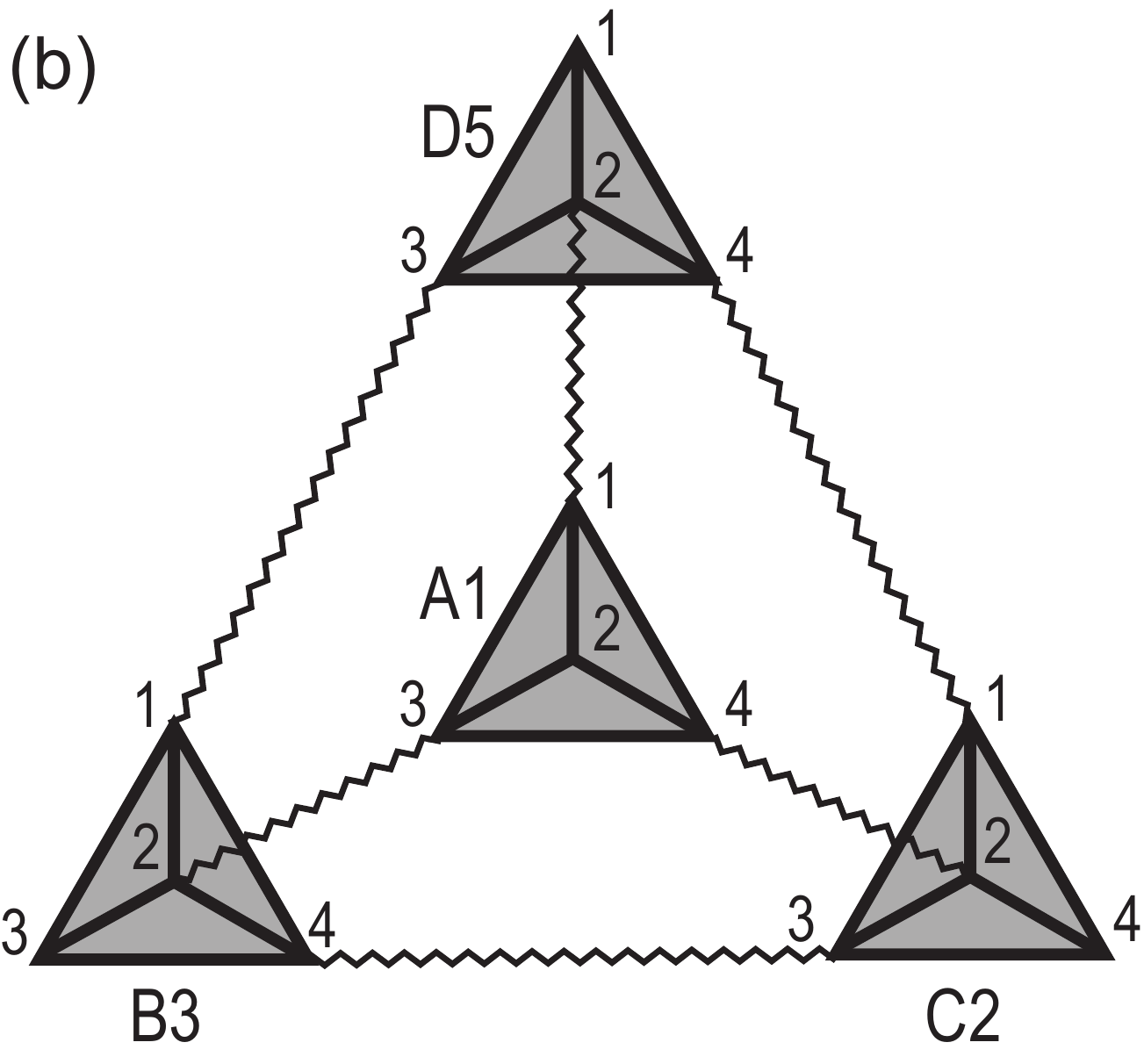}
\end{center}
\caption{
(a) Network of effective interactions.  
Since tetrahedron units 
form a face center cubic lattice,  
each unit is surrounded by 12 nearest neighbors.  
The number accompanying $A$-$D$ 
is the label of cubic unit cell.  
The central unit $A$1 is surrounded by 4 $D$-units in the 
same layer, and 4 $B$- and also 4 $C$-units in 
the layer either below or above.  
Long bonds connecting $A$ and $D$'s are also shown.  
Setting one of its vertices as an apex, each unit form a quartet 
of tetrahedron units together with three neighbors.  
Shown with zigzag lines is the case where the apex is 
the site 2 of the unit $A$1. 
This quartet has three hexagon loops that attach the $A$1 
unit, and each loop corresponds to one term in the 
effective Hamiltonian (\ref{eq3:eff}).  
Therefore, the Hamiltonian has 
12 terms containing $\bm{\tau}_{A1}$. 
(b) Connectivity of 4 units in the tetrahedron quartet 
selected in (a).  
The numbers 1-4 are site index in each unit.  
\label{fig:network}}
\end{figure}

The neutrality identity implies that only two of the operators 
$\mathsf{F}^{(1j)}$'s are independent in each tetrahedron unit, and 
I choose the following two 
\begin{equation}
\tau_1=\mathsf{F}^{(12)}, \ \ 
\tau_2={\textstyle \frac{1}{\sqrt{3}}} 
(\mathsf{F}^{(13)}-\mathsf{F}^{(14)}) . 
\end{equation}
I have checked that these operators transform as 
basis of the $E$-irrep upon operations in the $T_d$ point group, 
and 
\begin{equation}
\mathsf{F}^{(12)} = \mathbf{e}_0 \cdot \bm{\tau} , \ \ 
\mathsf{F}^{(13)} = \mathbf{e}_1 \cdot \bm{\tau} , \ \ 
\mathsf{F}^{(14)} = \mathbf{e}_2 \cdot \bm{\tau} , \ \ 
\label{eq:Ftau}
\end{equation}
with $\mathbf{e}_m = \mathbf{e}(\mbox{$\theta$=$2m\pi /3$})$ where 
$\mathbf{e}(\theta )=(\cos \theta , \sin \theta )$.   
As will be shown in Eq.~(\ref{eq:S12}) and Appendix \ref{sec:AF13}, 
all of the operators $\{ \mathsf{F}^{1j} \}$ have eigenvalues 
\begin{equation}
\lambda_l = - {\textstyle \frac{2}{3}} S(S+1) 
+ {\textstyle \frac{1}{2}} l(l+1) , \ \ 
\ ( l \in \{ 0, 1, \cdots , 2S \} ) . 
\label{eq:evF}
\end{equation}

The full effective Hamiltonian in the third order is obtained 
by repeating the same calculation for all the tetrahedron triads 
participating to the shortest hexagon loops 
in the breathing pyrochlore lattice.  
With these new operators $\bm{\tau}$, the effective Hamiltonian 
for the spin-$S$ Heisenberg model (\ref{eq2:model}) is represented as 
\begin{equation}
H_{\mathrm{eff}} 
= - J_{\mathrm{eff}} \!\!\!\!\!
\sum_{\langle \mathbf{r},\mathbf{r}',\mathbf{r}'' \rangle} \!\!
\bigl( c_0 - \mathbf{e}_{0} \cdot \bm{\tau}_{\mathbf{r}} \bigr)
\bigl( c_0 - \mathbf{e}_{1} \cdot \bm{\tau}_{\mathbf{r}'} \bigr)
\bigl( c_0 - \mathbf{e}_{2} \cdot \bm{\tau}_{\mathbf{r}''} \bigr) , 
\label{eq3:eff}
\end{equation}
where the parameters are 
\begin{equation}
J_{\mathrm{eff}}=\frac{J}{6} \left( \frac{J'}{J} \right)^3  > 0, 
\hspace{5mm}
c_0={\textstyle \frac{1}{3}} S(S+1).  
\label{eq:Jeff}
\end{equation}
Here the sum is taken over all the tetrahedron triads explained before. 
As explained in the caption of Fig.~\ref{fig:network}, the number 
of terms in $H_{\mathrm{eff}}$ is 
$12 \times \frac{1}{4}N \times \frac{1}{3}=N$, where $N$ is 
the number of original spins and the factor $\frac{1}{3}$ comes from 
the fact that each term of this three-unit interaction 
is counted three times. 
Corresponding to the choice of three units out of the four 
sublattices $A$-$D$, $H_{\mathrm{eff}}$ has 4 types of terms 
and the prefactors $\mathbf{e}_m$'s in Eq.~(\ref{eq3:eff}) should be 
chosen as follows depending on sublattices
\begin{equation}
\begin{array}{cccc}
 & \ \ \mathbf{e}_0 \ \ &\ \  \mathbf{e}_1 \ \ & \ \ \mathbf{e}_2  \ \ \\
\hline
\mbox{type } (1) & A & B & C \\ 
\phantom{\mbox{type }} (2) & B & A & D \\
\phantom{\mbox{type }} (3) & C & D & A \\
\phantom{\mbox{type }} (4) & D & C & B \\
\hline
\end{array}
\end{equation}
One should note that each $\mathbf{e}_m$ 
appears once and only once in each triple product term 
in Eq.~(\ref{eq3:eff}). 
The prefactor $\mathbf{e}_m$ is determined 
by how its tetrahedron unit is connected to the hexagon loop. 
It is $\mathbf{e}_0 $ if the unit is connected with the site pair 1-2 or 3-4, 
$\mathbf{e}_1 $ for 1-3 or 2-4, and $\mathbf{e}_2 $ for 1-4 or 2-3.  

This effective Hamiltonian derived for general $S$ is identical to 
the one obtained in the previous studies 
for the $S=\frac{1}{2}$ case \cite{Tsunetsugu1,Tsunetsugu2}. 
The only but essential difference is that $\bm{\tau}_\mathbf{r}$ 
now operates in the local singlet space which has the dimension $2S+1$.  
For $S=\frac{1}{2}$, $\bm{\tau}$ are a half of Pauli matrices 
and $c_0$=$\frac14$, and the Hamiltonian ~(\ref{eq3:eff}) 
reduces to the effective model in 
Refs.~\cite{Tsunetsugu1,Tsunetsugu2} up to 
the numerical factors.\footnote[5]{
References \cite{Tsunetsugu1,Tsunetsugu2} use chiral bases 
for wavefunctions,  
while real bases are used in this work.  
Tetrahedra $A$-$D$ are also named differently.  
Except these definitions, the two results are equivalent.
}
Expanding the triple products in $H_{\mathrm{eff}}$, 
it is again found that all the terms linear in $\bm{\tau}$ vanish 
for any $S$.  

Before proceeding to the next step, 
I briefly comment on the classical limit $S = \infty$, 
and explain that the present perturbative approach fails there.  
With increasing $S$ to infinity, 
while the quantum Hamiltonian converges to the classical Heisenberg model, 
the quantum ground state does not continuously 
evolve to the ground state of the classical model. 
The reason is the following.  

As explained at the beginning of this section, 
the use of the effective Hamiltonian is limited 
to the ground state and the low-energy sector 
of the original Heisenberg model 
where the excitation energy is smaller than 
the spin gap
$\varDelta E < \Delta_s = J \bar{\Delta}_s (\frac{J'}{J};S)$.  
In the $S=\infty$ limit, 
the classical Heisenberg model is defined with classical 
unit vectors $\mathbf{s}_i (\mathbf{r})$'s as 
$ H_{\mathrm{cl}} 
= J_{\mathrm{cl}} \sum_{\mathbf{r}, i < j} 
\mathbf{s}_i (\mathbf{r}) \cdot \mathbf{s}_j (\mathbf{r})$ 
$+ J'_{\mathrm{cl}} \sum_{\langle (i, \mathbf{r}), (j,\mathbf{r}') \rangle} 
 \mathbf{s}_i (\mathbf{r}) \cdot \mathbf{s}_j (\mathbf{r}')$.  
To converge to this upon increasing $S$ in the quantum Hamiltonian 
(\ref{eq2:modelA}), one needs to renormalize spin variables as 
$\mathbf{S}_i (\mathbf{r}) = S \mathbf{s}_i (\mathbf{r})$, 
and this requires a proper scaling of the exchange constants 
\begin{equation} 
J = \frac{J_{\mathrm{cl}}}{S^2}, \ \ 
J' = \frac{J'_{\mathrm{cl}}}{S^2} , 
\end{equation} 
where $J_{\mathrm{cl}}$ and $J'_{\mathrm{cl}}$ 
are constants independent of $S$.  
This immediately implies that 
$\varDelta E < \frac{J_{\mathrm{cl}}}{S^2} 
 \bar{\Delta}_s (\frac{J'_{\mathrm{cl}}}{J_{\mathrm{cl}}} ;S) 
 \rightarrow 0$ 
with $S \rightarrow \infty$, and 
the energy region of the effective model shrinks to zero 
in the classical limit.  
Thus, the low-energy region of the quantum Hamiltonian 
(\ref{eq2:modelA}) with finite $S$ is not continuously 
connected to that in the classical Heisenberg model. 
In particular, the $S=\infty$ limit is a singular point 
for the ground state: 
the spin rotation symmetry is not broken 
in the ground state for any finite $S$, 
but it is broken in the classical ground state.  
Therefore, it is impossible to formulate 
an expansion of $\frac{1}{S}$ type starting from the classical limit.
This contrasts with the case of magnetically ordered states, 
where the $\frac{1}{S}$-expansion correctly describes 
the ordered ground state and 
magnon excitations in the corresponding quantum system.  
The classical Heisenberg model on the pyrochlore lattice 
is itself exotic, and the ground state is 
thermodynamically degenerate \cite{Reimers,Benton}. 

\section{Spin-pair operators $\bm{\tau}$ for general $S$}
To analyze the effective Hamiltonian, one needs to know 
an explicit form of $\bm{\tau}$ operators, and this 
is another challenge for $S > \frac{1}{2}$.  
In this section, I am going to calculate the matrix elements of 
$\bm{\tau}$ in terms of the basis states 
$\{ \Phi_l \}$ defined in Eq.~(\ref{eq2:CG4}).  
It turns out useful to introduce uniform and staggered 
components of spin pair, 
\begin{equation} 
\mathbf{S}_{ij} \equiv \mathbf{S}_i + \mathbf{S}_j, \ \ 
\mathbf{N}_{ij} \equiv \mathbf{S}_i - \mathbf{S}_j,  \ \ 
\label{eq:SandN}
\end{equation}
and their ladder operators,  
$S_{ij}^{\pm} \equiv S_{ij}^x \pm i S_{ij}^y$ 
and $N_{ij}^{\pm} \equiv N_{ij}^x \pm i N_{ij}^y$. 

The spin-pair wavefunction 
$\phi^{(12)}(l,l_z )$ is an eigenvector 
of $\mathbf{S}_{12}^2$ with the eigenvalue $l(l+1)$ for 
any $l_z$.  
Therefore, this is also the case for our basis functions 
in the $S_{\mathrm{unit}}=0$ subspace of tetrahedron unit, and 
\begin{equation} 
l(l+1) \Phi_{l} = \mathbf{S}_{12}^2 \Phi_{l} 
= 2 \bigl[ S(S+1) + \mathbf{S}_{1} \cdot \mathbf{S}_{2} \bigr] \Phi_{l} . 
\label{eq:phi12}
\end{equation}
This immediately leads to the matrix elements 
of $\tau_1$
\begin{equation} 
\bigl(\tau_1 \bigr)_{l' l} = F_{l' l}^{(12)}  = 
\langle  \Phi_{l'} | {\textstyle \frac{1}{3}} S(S+1) 
+ \mathbf{S}_1 \cdot \mathbf{S}_2 | \Phi_{l} \rangle
= 
\bigl[ {\textstyle \frac{1}{2}} l(l+1) 
     - {\textstyle \frac{2}{3}} S(S+1) \bigr] \delta_{l' l} . 
\label{eq:S12}
\end{equation}
This matrix is diagonal and traceless, 
$\sum_{l=0}^{2S} ( \tau_1 )_{ll}=0$. 
The largest and smallest eigenvalues are 
$\frac{1}{3} S (4S+1)$ and 
$-\frac{2}{3} S (S+1)$, respectively.  

The calculation of 
$\tau_2 = \frac{1}{\sqrt{3}} (\mathsf{F}^{(13)} - \mathsf{F}^{(14)} ) 
= \frac{1}{\sqrt{3}} (2\mathsf{F}^{(13)} - \mathsf{F}^{(12)} )$ 
is more elaborate, 
since the operation of $\mathbf{S}_1 \cdot \mathbf{S}_3$ or 
$\mathbf{S}_1 \cdot \mathbf{S}_4$ hybridizes different $\Phi_l$'s 
and four-spin nature of the wavefunctions complicates its evaluation.  
As shown in Appendix \ref{sec:AF13}, $\mathsf{F}^{(13)}$ is 
related to $\mathsf{F}^{(12)}$ by a unitary transformation, 
but this needs an involved calculation 
of many 9$j$- or 6$j$-symbols.  
I have found a practical way of directly calculating 
$\tau_2$ for general $S$, and I explain this in the following.  

Matrix element $( \tau_2 )_{l' l}$ is given by the overlap integral 
between $\Phi_{l'}$ and 
$\tilde{\Psi}_{l} \equiv  2 ( \mathbf{S}_1 \cdot \mathbf{S}_3 
    -  \mathbf{S}_1 \cdot \mathbf{S}_4 )  \Phi_l$ 
multiplied by factor $\frac{1}{2\sqrt{3}}$. 
It is important to notice that 
$\tilde{\Psi}_l$ is in the subspace of $S_{\mathrm{unit}}=0$.  
This is because the relation 
$[ \mathbf{S}_{\mathrm{unit}}^2 , \mathbf{S}_1 \cdot \mathbf{S}_j ]=0$ 
leads to the eigenvalue equation 
\begin{equation} 
\mathbf{S}_{\mathrm{unit}}^2 \tilde{\Psi}_l
= 
2 \bigl( \mathbf{S}_1 \cdot \mathbf{S}_3 - 
       \mathbf{S}_1 \cdot \mathbf{S}_4 \bigr) \,  
\mathbf{S}_{\mathrm{unit}}^2 \Phi_l = 0 .  
\end{equation}
Next, I rewrite $\tilde{\Psi}_l$ to a symmetric 
form for simplifying further calculation.  
The definition gives 
$\tilde{\Psi}_l  = 
( \mathbf{S}_{12} + \mathbf{N}_{12} \bigr) \cdot 
\mathbf{N}_{34} \Phi_l$, 
and the conjugate-pair equivalence leads to another expression 
$\tilde{\Psi}_l = 
2 ( \mathbf{S}_2 \cdot \mathbf{S}_4 - \mathbf{S}_2 \cdot \mathbf{S}_3 ) 
\Phi_l 
= 
\bigl( -\mathbf{S}_{12} + \mathbf{N}_{12} \bigr) \cdot 
\mathbf{N}_{34} \Phi_l $. 
Averaging these two, one obtains a more symmetric form 
\begin{equation} 
\tilde{\Psi}_l 
= 
\mathbf{N}_{12} \cdot \mathbf{N}_{34}  \Phi_l , 
\label{eq:NNPhi}
\end{equation}
and I am going to calculate this.  

Now, let us examine more details of $\tilde{\Psi}_l$. 
It reads in terms of pair wavefunctions as 
\begin{align}
&\tilde{\Psi}_l = 
\sum_{l_z} C(l,l_z ) \Bigl\{ 
N_{12}^z \phi^{(12)} (l,l_z) \otimes N_{34}^z \phi^{(34)}(l,-l_z) 
\nonumber\\
&\hspace{1cm} + 
{\textstyle \frac{1}{2}} \Bigl[ 
N_{12}^+ \phi^{(12)} (l,l_z) \otimes N_{34}^- \phi^{(34)}(l,-l_z) 
+ 
N_{12}^- \phi^{(12)} (l,l_z) \otimes N_{34}^+ \phi^{(34)}(l,-l_z) \Bigr] \Bigr\} .
\label{eq:PsiNN}
\end{align}
The goal is to express this in terms of our singlet basis functions 
$\{ \Phi_{l'} \}$.  
I have not been able to find the formula of operating 
$\mathbf{N}$ in the literature, and so I need to derive it. 

I start with the part operated by $N^z$ operator.   
The definition of the pair wavefunction (\ref{eq2:2spinwf}) 
leads to 
\begin{equation} 
N_{12}^z \phi^{(12)} (l,l_z ) 
= \sum_m \, \Bigl[ (2m-l_z) 
\langle S,S,m,l_z-m | l,l_z \rangle \Bigr] \, 
|S,m \rangle_1 \otimes |S,l_z -m \rangle_2 . 
\end{equation} 
Among various recursion formulas of the CG coefficient, useful 
is the one that changes the composite angular 
momentum\footnote[6]{%
Equation (\ref{eq:CGNz}) used in the present work is a special case 
of Eq.~(C.20) in Ref.~\cite{Messiah}, but the result in the reference 
is erroneous. 
$f(x)$ there should be multiplied by factor 2.  
}
\cite{Messiah}, 
\begin{align}
(2m-l_z) 
\langle S,S,m,l_z-m | l,l_z \rangle 
= 
\sqrt{l^2 - l_z^2 } \, B_S (l) \, 
&\langle S,S,m,l_z-m | l-1,l_z \rangle 
\nonumber\\
+ 
\sqrt{(l+1)^2 - l_z^2 } \, B_S (l+1) \, 
&\langle S,S,m,l_z-m | l+1,l_z \rangle , 
\label{eq:CGNz}
\end{align}
with 
\begin{equation} 
B_S (l) \equiv \left[ 
\frac{(2S+1)^2 - l^2}{4l^2 - 1} \right]^{1/2}, 
\end{equation} 
and the projection $l_z$ does not change here.  
Since the coefficients 
on the right-hand side of Eq.~(\ref{eq:CGNz}) do not 
depend on $m$, this leads to the same formula 
for the pair wavefunction 
\begin{equation} 
N_{12}^z \phi^{(12)} (l,l_z ) 
= \sqrt{l^2 - l_z^2 } \, B_S (l) \, 
\phi^{(12)} (l-1,l_z ) 
+ \sqrt{(l+1)^2 - l_z^2 } \, B_S (l+1) \, 
\phi^{(12)} (l+1,l_z ) . 
\end{equation} 

Operation of the ladder operators $N_{12}^{\pm}$ is 
more difficult to perform.  
Instead of their direct operation, it is useful to notice the following 
identity 
\begin{equation} 
N_{12}^{\pm} = \pm  \bigl[ N_{12}^z , S_{12}^{\pm} ] . 
\end{equation} 
and operate these commutators instead.  
In this case, one knows all the necessary matrix elements.  
Operating $S_{12}^{\pm}$ changes $l_z$ by $\pm 1$, 
and $N_{12}^z$ changes $l$ by $\pm 1$.  
The result is 
\begin{eqnarray} 
N_{12}^{\pm} \phi^{(12)} (l,l_z ) 
= \pm \sqrt{(l \mp l_z -1)(l \mp l_z)  } \ B_S (l) \ 
\phi^{(12)} (l-1,l_z \pm 1 ) \phantom{.} &&
\nonumber\\
\mp \sqrt{(l \pm l_z +1)(l \pm l_z +2)  } \ B_S (l+1) \ 
\phi^{(12)} (l+1,l_z \pm 1) .  && 
\end{eqnarray}
With these results, we go back to Eq.~(\ref{eq:PsiNN}) and 
sum over $l_z$ on the right-hand side.  
This summation contains two types of products concerning 
the composite spin: 
$\phi^{(12)}(l+\varDelta l,\cdot ) 
 \otimes \phi^{(34)}(l+\varDelta l,\cdot )$, and 
$\phi^{(12)}(l+\varDelta l,\cdot ) 
\otimes \phi^{(34)}(l-\varDelta l,\cdot )$, where 
$\varDelta l = \pm 1$. 
Those of the latter type do not belong to the 
subspace of $S_{\mathrm{unit}}=0$, since the two 
composite spins differ.  
As proved before, $\tilde{\Psi}_l$ 
is a wavefunction in the $S_{\mathrm{unit}}=0$ subspace, 
and therefore these cross terms cancel to each other.  
The products of the former type contribute to the singlet components 
$\Phi_{l-1}$ and $\Phi_{l+1}$ as 
$
 \tilde{\Psi}_l 
= - 2 \sqrt{3} [ 
 f_S (l ) \Phi_{l-1} + f_S (l+1) \Phi_{l+1} ], 
$
with the coefficient 
\begin{equation} 
f_S ( l ) \equiv \, \frac{l}{2 \sqrt{3}} \, 
\frac{(2S+1)^2 - l^2}{\sqrt{4l^2 - 1}} . 
\end{equation}
This completes the calculation of $\tau_2$. 
The matrix elements are given by 
\begin{equation} 
\bigl( \tau_2 \bigr)_{l' l} 
= 
-  f_S (l ) \, \delta_{l',l-1} 
-  f_S (l+1) \, \delta_{l',l+1}  , 
\label{eq:tau2S}
\end{equation} 
and this matrix is tridiagonal with zero diagonal elements.

\section{Mean field theory of the effective model}
\label{sec:MFgs}

\subsection{Mean field equation}
The final step is the task of solving the effective Hamiltonian 
$H_{\mathrm{eff}}$. 
I do this by a mean field approximation at zero temperature.  
First, I examine in this section this problem for general $S$, 
and later in the following sections obtain explicit solutions for 
the cases of $S=\frac{3}{2}$ and 1 and discuss the results in detail. 
This approach is equivalent to approximating the ground state by a product of 
local wavefunctions of all the tetrahedron units and those 
local wavefunctions are to be optimized.  
I further assume that 
the spatial pattern has the cubic unit cell with 16 original spins, 
corresponding to the four tetrahedron units $A$-$D$ in Fig.~\ref{fig:lattice}, 
and the translation symmetry is not broken further.  
In this case, a trial product state reads as 
\begin{equation}
\Psi_{\mathrm{trial}} 
= \bigotimes_{\mathbf{R}} 
\Bigl[         \psi_A (\mathbf{r}_A + \mathbf{R}) 
       \otimes \psi_B (\mathbf{r}_B + \mathbf{R}) 
       \otimes \psi_C (\mathbf{r}_C + \mathbf{R}) 
       \otimes \psi_D (\mathbf{r}_D + \mathbf{R}) \Bigr] , 
\label{eq:MFwf2}
\end{equation}
where $\mathbf{R}$ denotes the position of cubic unit cell.  
Each $\psi_X$ is a (2$S$+1)-dimensional trial wavefunction 
in the unit $X$, and $\{ \psi_A , \cdots , \psi_D \}$ are 
to be determined by energy minimization.  
Equivalently, one may define the ``order parameters'' by 
a set of 4 two-dimensional real vectors 
$\{ \langle \bm{\tau}_X \rangle \}_{X=A}^D$, where 
$\langle \bm{\tau}_X \rangle = \langle \psi_X | \bm{\tau} | \psi_X \rangle$, 
and determine them by minimizing the mean field energy 
\begin{align}
E_{\mathrm{MF}} = 
{\textstyle \frac{16}{N}} 
\bigl\langle 
\Psi_{\mathrm{trial}} \big| H_{\mathrm{eff}} \big| \Psi_{\mathrm{trial}} 
\bigr\rangle = 
&4J_{\mathrm{eff}}
\bigl( 1 + P_{AB}P_{CD} + P_{AC}P_{BD}+ P_{AD}P_{BC} \bigr) 
\nonumber\\
&\hspace{0cm} \times 
\bigl( -c_0  + \mathbf{e}_0
       \cdot \langle \bm{\tau}_A \rangle\bigr)
\bigl( -c_0  + \mathbf{e}_1 
       \cdot \langle \bm{\tau}_B \rangle\bigr)
\bigl( -c_0  + \mathbf{e}_2 
       \cdot \langle \bm{\tau}_C \rangle\bigr) . 
\label{eq:EMF}
\end{align}
where $c_0 = \frac{5}{4}$ in the $S=\frac{3}{2}$ case 
and $\frac{1}{3} S(S+1)$ for general $S$.  
$\mathbf{e}_m = 
(\cos \frac{2m\pi}{3}, \sin \frac{2m\pi}{3})$ 
as before 
and $P_{XY}$ denotes the operation that exchanges the units $X$ and $Y$.  
Beware that $\langle \bm{\tau}_X \rangle$ is related to 
two-spin correlation in the unit $X$ 
\begin{equation}
\langle \mathbf{S}_1 \cdot \mathbf{S}_j \rangle_X 
= - c_0 
  + \mathbf{e}_{j-2}  \cdot \langle \bm{\tau}_X \rangle . \ \ 
(j \in \{2,3,4 \}) 
\label{eq:twospin}
\end{equation}
This relation holds for general $S$.  

Minimizing $E_{\mathrm{MF}}$ with respect to $\psi_X$ reduces to 
an eigenvalue problem,
$H_{\mathrm{MF}}^X \psi_X = \epsilon_X \psi_X$, 
with the mean field Hamiltonian 
\begin{equation}
H_{\mathrm{MF}}^X / (4J_{\mathrm{eff}}) = - \mathbf{h}_X \cdot \bm{\tau}_X , 
\label{eq:mfHX}
\end{equation}
and the mean field at the unit $A$ is given by 
\begin{eqnarray}
&&\mathbf{h}_A = 
-\mathbf{e}_0 
\bigl(-c_0  + \mathbf{e}_1 \!
       \cdot \! \langle \bm{\tau}_B \rangle\bigr)
\bigl(-c_0  + \mathbf{e}_2 \!
       \cdot \! \langle \bm{\tau}_C \rangle\bigr)
-\mathbf{e}_1
\bigl(-c_0  + \mathbf{e}_2 \!
       \cdot \! \langle \bm{\tau}_D \rangle\bigr)
\bigl(-c_0  + \mathbf{e}_0 \!
       \cdot \! \langle \bm{\tau}_B \rangle\bigr)
\nonumber\\
&&\hspace{1cm} 
-\mathbf{e}_2
\bigl(-c_0  + \mathbf{e}_0 \!
       \cdot \! \langle \bm{\tau}_C \rangle\bigr)
\bigl(-c_0  + \mathbf{e}_1 \!
       \cdot \! \langle \bm{\tau}_D \rangle\bigr) . 
\label{eq:hA}
\end{eqnarray}
Similar results are also obtained for the other units $BCD$.  
One needs to determine 
$\langle \bm{\tau}_A \rangle, \cdots, \langle \bm{\tau}_D \rangle$ 
self-consistently.

\subsection{Single-unit problem} 
\label{subsec:1unit}
To solve the self-consistent equations in the mean field approach, 
one needs to calculate 
$\langle \bm{\tau}_X \rangle$ for a given mean field field $\mathbf{h}_X$, 
and I now discuss this problem in more detail.  
Calculations up to this stage are common for general $S$. 
However, solutions of the single unit problem have different 
characters depending on the value of $S$, and 
the case of $S=\frac{1}{2}$ is exceptional as I will show below.  

Eigenstates of $H_{\mathrm{MF}}^X$ are completely 
determined by the direction $\zeta$ of the mean field, 
$\mathbf{h}_X /|\mathbf{h}_X | \equiv \mathbf{e}(\zeta )$. 
Therefore, it is sufficient to define the dimensionless Hamiltonian 
and consider its ground state $\psi_0 (\zeta )$: 
\begin{equation}
\bar{H}_{\mathrm{MF}}(\zeta ) = - \mathbf{e}(\zeta ) \cdot \bm{\tau} , \ \ 
\bar{H}_{\mathrm{MF}}(\zeta ) \psi_0 (\zeta ) 
= \epsilon_0 (\zeta ) \psi_0 (\zeta ) .  
\label{eq:HMF1}
\end{equation}
Here $\epsilon_0 (\zeta ) $ is the dimensionless ground-state 
energy and this generally depends on the field direction $\zeta$.  
The only exception is the case of $S=\frac{1}{2}$, where 
$\tau_1$ and $\tau_2$ are both half Pauli matrix and therefore 
$\epsilon_0 (\zeta ) = - \frac{1}{2}$ for any $\zeta$.  

Regarding the field direction dependence, 
it is important to notice that as will be proved later, 
the $T_d$ point group of the tetrahedron unit implies the following 
properties of the ground state energy $\epsilon_0 (\zeta )$ for any $S$  
\begin{equation} 
\epsilon_0 \bigl( \zeta \pm {\textstyle \frac{2\pi}{3}} ) 
= \epsilon_0 (\zeta ) , \ \ 
\epsilon_0 ( - \zeta  ) = \epsilon_0 (\zeta ) . \ \ 
\label{eq:epsilon_periodic}
\end{equation}
This shows that the field anisotropy has the $Z_3$ symmetry, and 
also means that $\epsilon_0 (\zeta )$ is extreme for the 
6 directions 
\begin{equation} 
\epsilon_0 ' (\zeta ) = 0 , \ \ \ 
\mbox{at } \zeta = {\textstyle \frac{\pi}{3}} \times (\mbox{integer}) 
\end{equation} 
where $'$ denotes the derivative with respect to $\zeta$.  
Since $\bar{H}_{MF} (0) = -\tau_1$ and $\bar{H}_{MF} (\pi ) = + \tau_1$, 
the extreme values of $\epsilon_0 (\zeta )$ are given by $\tau_1$'s 
largest and smallest eigenvalues 
obtained in Eq.~(\ref{eq:S12}) 
\begin{equation}
\epsilon_0^{\mathrm{min}} = 
\epsilon_0 (0 ) = -{\textstyle \frac{1}{3}} S (4S+1), \ \ \ 
\epsilon_0^{\mathrm{max}} = 
\epsilon_0 (\pi ) = -{\textstyle \frac{2}{3}} S (S+1) . 
\end{equation} 
Note that the above arguments claim only local extremeness 
and do not guarantee these are global maximum or minimum 
in the whole $\zeta$ region.  
However, calculations for $S=\frac{3}{2}$ and 1 show that they are 
the global maximum and minimum, and this suggests this also holds 
for general $S$.   

The strength of the $Z_3$ anisotropy is characterized by the ratio 
of the minimum and maximum values of the ground state energy
\begin{equation} 
R_{\mathrm{anis}} \equiv 
\frac{\ \epsilon_0^{\mathrm{min}}}{\ \epsilon_0^{\mathrm{max}}}
= 2 - \frac{3}{2S+2} \ge 1 . 
\label{eq:Ranis}
\end{equation} 
For $S=\frac{1}{2}$, this parameter reduces to $R_{\mathrm{anis}}=1$, 
and the anisotropy completely vanishes.  
For larger $S$, the anisotropy monotonically increases with $S$ 
and approaches 2 in the $S=\infty$ limit.  
One should note that this anisotropy parameter also 
gives the ratio of the maximum and minimum moduli of the order 
parameter vector 
$R_{\mathrm{anis}} = 
\max_{\zeta} |\langle \bm{\tau}  (\zeta ) \rangle| / 
\min_\zeta |\langle \bm{\tau } (\zeta ) \rangle |
= 
|\langle \bm{\tau}  (0 ) \rangle| / 
|\langle \bm{\tau } (\pi ) \rangle |$.  

Now, I quickly prove the symmetries (\ref{eq:epsilon_periodic}). 
The first and second equalities are related to 
the two symmetry operations $P_5$ and $P_4$ in the tetrahedron unit, 
respectively, introduced in Appendix \ref{sec:A1}.  
$P_5$ is a 3-fold rotation about the site 1, while 
$P_4$ is a diagonal mirror operation that exchanges the sites 1 and 2.  
These operations transform $\bm{\tau}$ operators as follows 
\begin{subequations}
\begin{align}
&P_5 \, \tau_1 \, {}^{t} \! P_5 
= P_5 \, \mathsf{F}^{(12)} \, {}^{t} \! P_5 
= \mathsf{F}^{(13)} 
= - {\textstyle \frac{1}{2}} \, \tau_1 
+ {\textstyle \frac{\sqrt{3}}{2}} \, \tau_2 , 
\label{eq:PtP1}
\\
&P_5 \, \tau_2 \, {}^{t} \! P_5 
= P_5 \, {\textstyle \frac{1}{\sqrt{3}}} 
\bigl( \mathsf{F}^{(13)} - \mathsf{F}^{(14)} \bigr) \, {}^{t} \! P_5 
= {\textstyle \frac{1}{\sqrt{3}}} 
\bigl( \mathsf{F}^{(14)} - \mathsf{F}^{(12)} \bigr) 
= - {\textstyle \frac{\sqrt{3}}{2}} \, \tau_1 
- {\textstyle \frac{1}{2}} \, \tau_2 , 
\label{eq:PtP2}
\\
&P_4 \, \tau_1 \, {}^{t} \! P_4 = \tau_1 , \ \ 
P_4 \, \tau_2 \, {}^{t} \! P_4 = - \tau_2 , 
\label{eq:tau_transf}
\end{align}
\end{subequations}
where ${}^t \! P_n = P_n^{-1}$ and 
Eq.~(\ref{eqA:F234}) is used for the first two relations.  
For the relations with $P_4$, I have used 
$(P_4 )_{l' l} = (-1)^{2S+l} \delta_{l' l}$ as well as 
Eqs.~(\ref{eq:S12}) and (\ref{eq:tau2S}).  
Note that these operations are equivalent to 
rotation and mirror in the $\bm{\tau}$-space
\begin{equation} 
P_4 \, \bm{\tau} \, {}^{t} \! P_4 
= \left( 
\begin{array}{cc} 
1 & 0 \\
0 & -1 
\end{array} 
\right) \bm{\tau} , 
\hspace{0.4cm}
P_5 \, \bm{\tau} \, {}^{t} \! P_5 
= {}^t R \bigl( {\textstyle \frac{2}{3}\pi} \bigr) 
\bm{\tau} , 
\hspace{0.4cm} 
\mbox{where }
R (\theta ) \equiv 
\left( 
\begin{array}{cc} 
\cos \theta & -\sin \theta \\
\sin \theta & \cos \theta 
\end{array} 
\right) . 
\label{eq:PPtau}
\end{equation}
Using these relations, it is straightforward to show 
the following transformation for the Hamiltonian 
\begin{equation} 
P_5 \, \bar{H}_{\mathrm{MF}} (\zeta ) \, {}^{t} \! P_5 
= \bar{H}_{\mathrm{MF}} \bigl(\zeta + {\textstyle \frac{2\pi}{3}} \bigr) , \ \ 
P_4 \, \bar{H}_{\mathrm{MF}} (\zeta ) \, {}^{t} \! P_4 = 
\bar{H}_{\mathrm{MF}} (- \zeta ) . 
\end{equation} 
Since $P_5$ and $P_4$ are both orthogonal matrix, this guarantees that 
the 6 cases of $\pm \zeta$, $\pm \zeta + \frac{2\pi}{3}$, and 
$\pm \zeta - \frac{2\pi}{3}$ have the identical energy spectrum, 
and the properties (\ref{eq:epsilon_periodic}) are proved.  
The ground state wavefunctions are related as 
\begin{equation} 
\psi_0 \bigl(\zeta + {\textstyle \frac{2\pi}{3}} \bigr) 
= P_5 \psi_0 (\zeta ) , \ \ 
\psi_0 (- \zeta ) = P_4 \psi_0 (\zeta ) . 
\end{equation}

Let me also discuss the order parameter defined 
for the ground state, 
$\langle \bm{\tau} (\zeta ) \rangle \equiv$
$\langle \psi_0 (\zeta ) | \bm{\tau} |\psi_0 (\zeta ) \rangle$.  
This expectation value is also transformed as shown in Eq.~(\ref{eq:PPtau}) 
with symmetry operations, and therefore manifests 
symmetry breaking of the $T_d$ point group.  
Once the ground-state energy is obtained, one can calculate this 
without using $\psi_0 (\zeta )$.  
To this end, 
the Hellmann-Feynman theorem \cite{Hellmann} 
is useful and this yields the result 
\begin{equation}
\langle \bm{\tau} (\zeta ) \rangle = 
-\epsilon_0 (\zeta ) \mathbf{e}(\zeta ) -
\epsilon_0 ' (\zeta ) \mathbf{e} '(\zeta ), 
\label{eq:OPzeta}
\end{equation}
where $'$ again denotes the derivative with respect to $\zeta$.  
This means that the component parallel to the applied field 
has amplitude $-\epsilon_0 (\zeta )$.  
The transverse component has amplitude $-\epsilon_0 ' (\zeta )$, and thus does 
not vanish unless the mean field points to any of the symmetric 
directions  
$\zeta = \frac{1}{3} \pi$$\times$(integer).  

\begin{figure}[bt]
\begin{center}
\includegraphics[height=7cm]{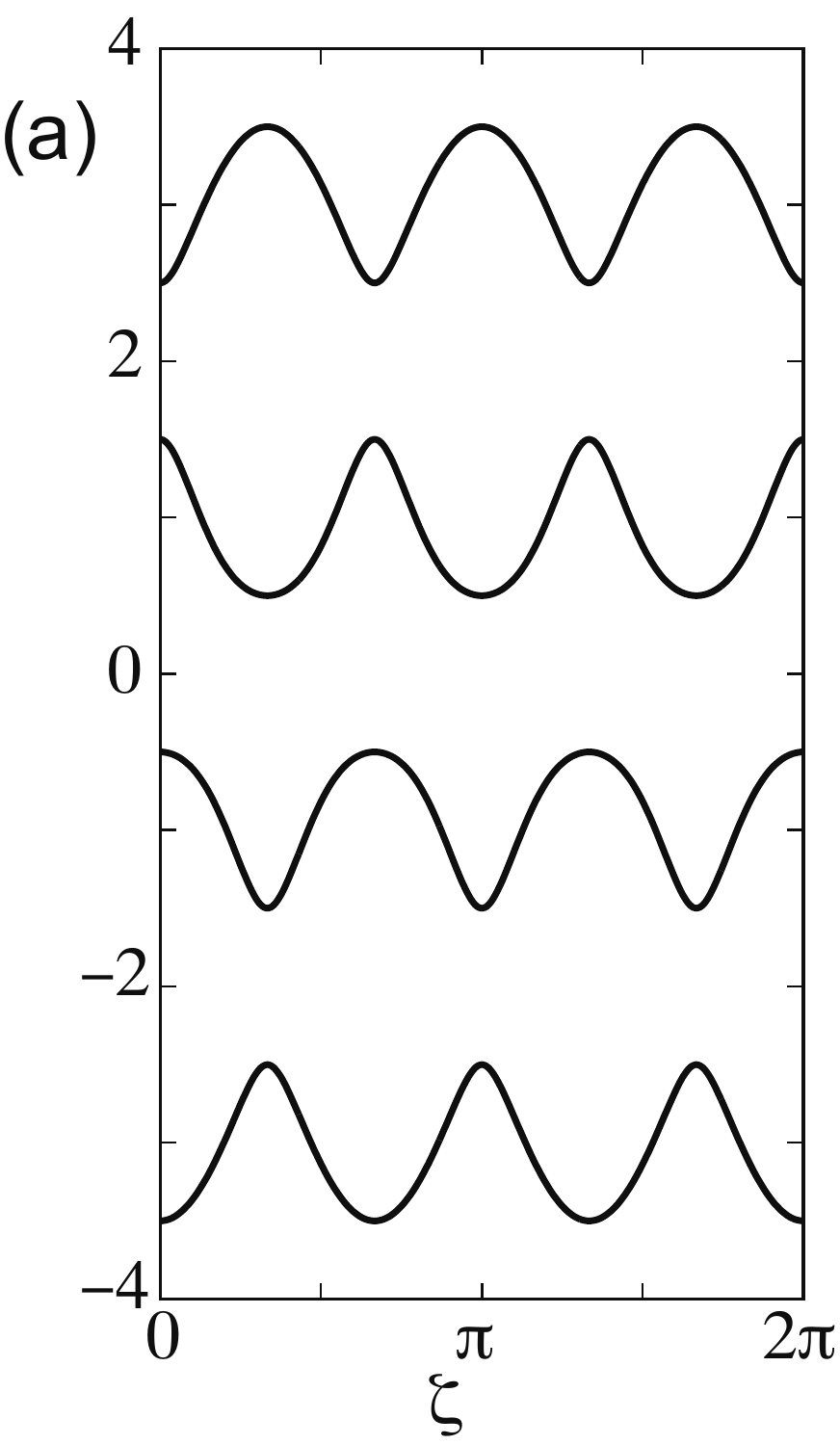}
\hspace{2cm}
\includegraphics[height=7cm]{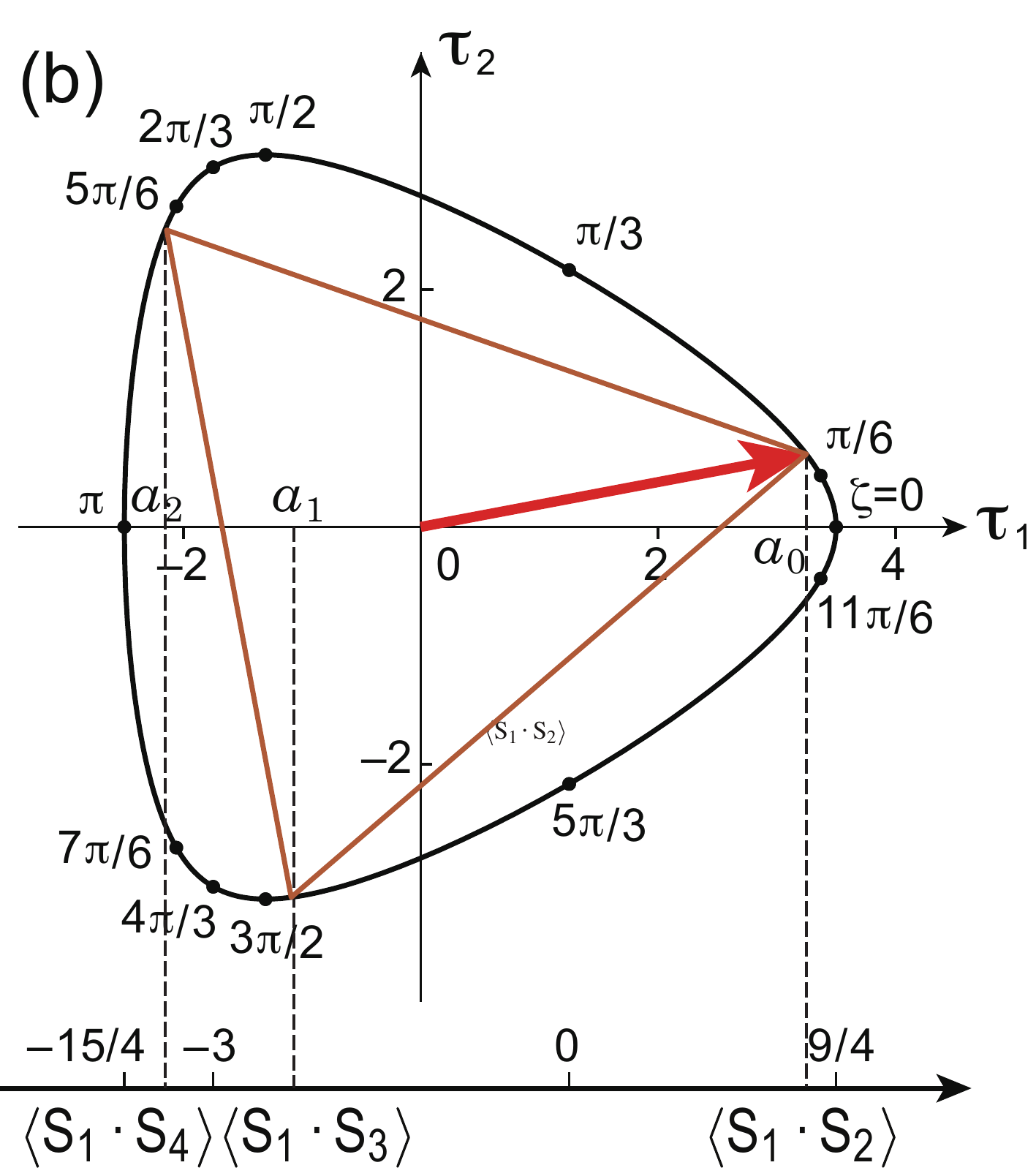}
\end{center}
\caption{(a) Eigenenergies of 
$\bar{H}_{\mathrm{MF}}(\zeta )$ for the $S=\frac32$ case.  
(b) Trajectory of $\langle \bm{\tau} \rangle$ calculated for the ground state 
of $\bar{H}_{\mathrm{MF}}(\zeta )$ is a rounded triangle.  
Spin correlations 
$\langle \mathbf{S}_1 \cdot \mathbf{S}_{2+m} \rangle + \frac{5}{4}$
=$\mathbf{e}_m \cdot \langle \bm{\tau} \rangle \equiv a_m $ 
($m=0,1,2$) are projections of the three vertices of 
the equilateral triangle constructed from the vector 
$\langle \bm{\tau} \rangle$. 
For any wavefunction, its expectation value $\langle \bm{\tau} \rangle$ 
is located on this trajectory or in its interior region. 
\label{fig:tau}}
\end{figure}

\section{Mean-field ground state of the $S=\frac{3}{2}$ case}
I now begin investigating specific cases and start from the case 
of $S=\frac{3}{2}$, which is 
relevant for the compound Li$X$Cr$_4$O$_8$\cite{Okamoto1,Tanaka,Okamoto2}, and 
will obtain the mean-field ground state of the effective Hamiltonian.  

In this case, the $S_{\mathrm{unit}}=0$ space has dimension 4 
and its basis states belong to the $A_1$-, $A_2$-, and E-irreps 
of the $T_d$ point group 
\begin{eqnarray}
\Phi_{A_1} = \bigl( \sqrt7 \Phi_{1} - \sqrt3 \Phi_{3} \bigr)/\sqrt{10} ,
\hspace{0.2cm} 
&&\Phi_{A_2} = \bigl( \Phi_{0} - \sqrt5 \Phi_{2} \bigr)/\sqrt{6} , 
\nonumber\\
\Phi_{Eu} = \bigl( \sqrt3 \Phi_{1} + \sqrt7 \Phi_{3} \bigr)/\sqrt{10}, 
\hspace{0.2cm} 
&&\Phi_{Ev} = - \bigl( \sqrt5 \Phi_{0} + \Phi_{2} \bigr)/\sqrt{6}. 
\end{eqnarray}
Here, $\{ \Phi_{l} \}_{l=0}^3$ were defined in Eq.~(\ref{eq2:CG4}). 

\subsection{$\bm{\tau}$ operators in the $S=\frac{3}{2}$ case 
and solution of a single unit problem}
\label{sec:tau32}

With this basis set 
$\{ \Phi_{A_1}, \Phi_{A_2}, \Phi_{Eu}, \Phi_{Ev} \}$, 
it is straightforward to represent $\tau_1$ and $\tau_2$.  
The results read 
\begin{equation}
\tau_1 = \frac{1}{2} \left[ \begin{array}{cccc} 
0           &  0 & -\sqrt{21} & 0 \\
0           &  0 &  0 & \sqrt5 \\
-\sqrt{21}  &  0 &  4 & 0 \\
0   & \sqrt{5} & 0 & -4 
\end{array} \right], \ \ 
\tau_2 = \frac{1}{2} \left[ \begin{array}{cccc} 
0           &  0 & 0 & \sqrt{21}  \\
0           &  0  & \sqrt5 & 0 \\
0 & \sqrt{5}  &  0 & 4 \\
\sqrt{21} & 0 & 4 & 0 
\end{array} \right]. \ \ 
\end{equation}
One important point is that they have no matrix elements 
in the subspace spanned by $\Phi_{A_1}$ and $\Phi_{A_2}$. 
This is a consequence of the fact that $\bm{\tau}$ operators 
transform as bases of the $E$-irrep, since 
the product representations $A_1 \otimes E$ and $A_2 \otimes E$ 
contain neither $A_1$ or $A_2$ irrep.  

The result above shows an interesting difference between these 
two operators. 
To see this, let us divide the local $S_{\mathrm{unit}}=0$ space 
to two subspaces $V_{+}$ and $V_{-}$ that are spanned by 
$\{ \Phi_{A_1}, \Phi_{Eu} \}$ and $\{ \Phi_{A_2}, \Phi_{Ev} \}$, respectively. 
I should note that 
$V_-$ is the space of wavefunctions which change sign upon 
exchange of spins 1 and 2 
(i.e., permutation $P_4$ in Appendix \ref{sec:A1}) or 
equivalently 3 and 4, 
$\Phi \in V_{-} \rightarrow  P_{4} \Phi = - \Phi $, while 
$\Phi \in V_{+} \rightarrow  P_{4} \Phi = + \Phi $. 
Then, $\tau_1$ has no finite matrix elements between $V_+$ and $V_-$, 
while $\tau_2$'s finite matrix elements are only between them. 
This is another manifestation of the transformation (\ref{eq:tau_transf}). 
 
The eigenvalues of $\tau_1$ are 
$-\frac{5}{2}, -\frac{3}{2}, \frac{1}{2}, \frac{7}{2}$, 
while $\tau_2$'s eigenvalues are 
$\pm \bigl( 21/4 + \sqrt{21} \bigr)^{1/2}$ and 
$\pm \bigl( 21/4 - \sqrt{21} \bigr)^{1/2}$.  
$\tau_1$'s eigenvalues $-\frac{5}{2}$ and $\frac{1}{2}$ have 
eigenvectors in $V_-$, while 
$-\frac{3}{2}$ and $\frac{7}{2}$ have eigenvectors in $V_+$.  

The mean-field Hamiltonian $\bar{H}_{\mathrm{MF}} (\zeta )$ 
is now explicitly represented by a $4 \times 4$ matrix.  
I have diagonalized it and 
found that its four eigenenergies are all non-degenerate 
for any direction $\zeta$ as shown in Fig.~\ref{fig:tau}(a).  
The lowest eigenvalue is 
\begin{equation}
\epsilon_0 (\zeta )=
- g(\zeta ) - \sqrt{\frac{21}{4} -g(\zeta )^2 
+ \frac{2}{g(\zeta )} \cos 3 \zeta \ } , 
\label{eq:e0zeta}
\end{equation} 
where $g(\zeta)$ is a positive parameter given by 
\begin{equation}
g(\zeta )^2=\frac{7}{4} + \frac{\sqrt{21}}{2}
 \cos \left[ \frac{1}{3} \cos^{-1} \left(
 \frac{49+32 \cos^2 3\zeta }{21^{3/2}}
\right) \right] . 
\end{equation} 
and 
$\bigl[  \frac{21}{8} \bigl( 1 + \sqrt{\frac{5}{21}} \bigr) \bigr]^{1/2}
\approx 1.9763 \le g(\zeta ) \le 2$.  
As shown in Fig.~\ref{fig:tau}(a), $\epsilon_0 (\zeta )$ is minimum 
for the three field directions $\zeta= 2m\pi /3$ ($m$=0, 1, 2)
while maximum for $\zeta= (2m+1) \pi /3$.  
The order parameter $\langle \bm{\tau} (\zeta )\rangle$ is calculated 
from the formula (\ref{eq:OPzeta}) and 
its trajectory is plotted in Fig.~\ref{fig:tau}(b).  
This implies that 
the size of the local order parameter is limited as 
$| \langle \bm{\tau} \rangle | \le \frac{7}{2}$ for any state 
in the $S_{\mathrm{unit}}=0$ subspace, 
and $-\frac{5}{2} \le 
\mathbf{e}_m \cdot \langle \bm{\tau} \rangle \le \frac{7}{2}$, 
which is equivalent to 
$-\frac{15}{4} = -S(S+1)\le 
\langle \mathbf{S}_1 \cdot \mathbf{S}_j \rangle \le \frac{9}{4} = S^2$. 

The neutrality identity (\ref{eq:Neutral}) imposes a further 
constraint on the three $\langle \mathbf{S}_1 \cdot \mathbf{S}_j \rangle$'s.  
Since their sum $-S(S+1)$ agrees with the lower bound of 
$\langle \mathbf{S}_1 \cdot \mathbf{S}_j \rangle$, 
the partial sum of any two correlations is bounded from above 
\begin{equation} 
\langle \mathbf{S}_1 \cdot \mathbf{S}_{j_1} \rangle 
+ \langle \mathbf{S}_1 \cdot \mathbf{S}_{j_2} \rangle 
= -S(S+1) - \langle \mathbf{S}_1 \cdot \mathbf{S}_{j_3} \rangle 
\le 
-S(S+1) + S(S+1) =0  , \ \ 
\end{equation} 
where three $j$'s are all different.  
This manifests no possibility of two ferromagnetic 
spin pairs in any tetrahedron unit.  
Almost always, just two pairs should be antiferromagnetic 
and the remaining one should be ferromagnetic in the ground state.  
The only exception is the case of a pair of spin-singlet dimers: 
one $\langle \mathbf{S}_1 \cdot \mathbf{S}_j \rangle$  is antiferromagnetic 
and the other two are zero. 
As far as $| \langle \bm{\tau} (\zeta ) \rangle |$ is minimum 
at $\zeta =\pi$, 
this result holds for general $S$, because the neutral identity 
and the $\tau_1$'s lower bound are common for all $S$. 
However, beyond the mean-field approximation or at finite temperature, 
which I do not discuss in the present work, 
$\langle \bm{\tau} \rangle$ shrinks to a point in the interior region 
in Fig.~\ref{fig:tau}(b), and 
it is also possible that all the three spin pairs are antiferromagnetic, 
if the shrinking is large.   

\begin{figure}[b]
\begin{center}
\includegraphics[width=11cm]{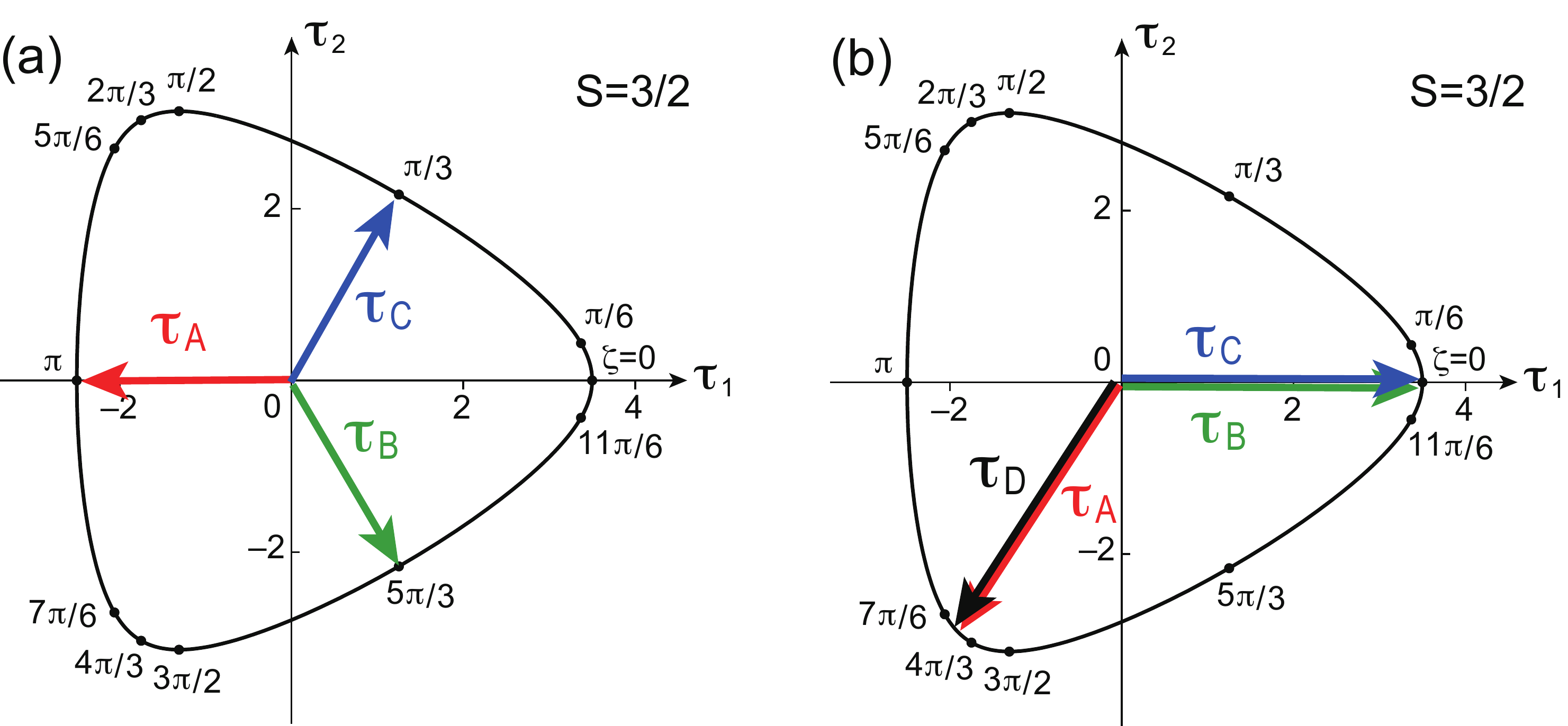}
\end{center}
\caption{
(a) The unique mean-field ground state for the triad of 
tetrahedron units $ABC$.  
(b) One of the mean-field ground states 
for the four units $ABCD$. 
This is also a solution in the bulk. 
\label{fig:mfsol}}
\end{figure}

\subsection{Solution for a triad of tetrahedron units}
\label{subsec:MFsol_triad}
With this result, let us to find 
the lowest-energy solution for one tetrahedron triad, e.g. $ABC$.  
The corresponding mean field energy is
\begin{equation}
E_{\mathrm{MF}} (ABC) =  - 4J_{\mathrm{eff}}
\Bigl({\textstyle \frac54} - \mathbf{e}_0
       \cdot \langle \bm{\tau}_A \rangle\Bigr)
\Bigl({\textstyle \frac54} - \mathbf{e}_1 
       \cdot \langle \bm{\tau}_B \rangle\Bigr)
\Bigl({\textstyle \frac54} - \mathbf{e}_2 
       \cdot \langle \bm{\tau}_C \rangle\Bigr)  , 
\label{eq:EMF_ABC}
\end{equation}
and the mean field is, for example, 
$\mathbf{h}_A = - \mathbf{e}_0  
\bigl({\textstyle \frac54} - \mathbf{e}_1 
       \cdot \langle \bm{\tau}_B \rangle\bigr)
\bigl({\textstyle \frac54} - \mathbf{e}_2 
       \cdot \langle \bm{\tau}_C \rangle\bigr)$.  
The lowest-energy solution is the one in which 
$\mathbf{e} (\theta_X) \cdot \langle \bm{\tau}_X \rangle$=$-\frac52$ 
for all the units $X$'s and its energy is 
$E_{\mathrm{MF}} (ABC) /(4 J_{\mathrm{eff}}) =-(15/4)^3 \approx$$-52.73$.  
This means that the order parameters are 
$\langle \bm{\tau} (\zeta) \rangle$ 
with $\zeta$=$\pi$, $\frac53 \pi$, and $\frac{1}{3} \pi$ for 
$\theta_X$=0, $\frac23 \pi$, and $\frac43 \pi$. 
Thus, the mean field points towards the direction 
opposite to $\mathbf{e} (\theta_X )$ at all the units, 
and the three $\langle \bm{\tau} \rangle$'s form 
an equilateral triangle. 
See Fig.~\ref{fig:mfsol}(a).  
This solution has interesting spin correlations.  
In each tetrahedron unit, only one pair of bonds 
have a strong antiferromagnetic correlation 
\begin{equation} 
\langle \mathbf{S}_1 \cdot \mathbf{S}_2 \rangle_A 
= \langle \mathbf{S}_3 \cdot \mathbf{S}_4 \rangle_A
= \langle \mathbf{S}_1 \cdot \mathbf{S}_3 \rangle_B
= \langle \mathbf{S}_2 \cdot \mathbf{S}_4 \rangle_B
= \langle \mathbf{S}_1 \cdot \mathbf{S}_4 \rangle_C 
= \langle \mathbf{S}_2 \cdot \mathbf{S}_3 \rangle_C 
= - {\textstyle \frac{15}{4}} , 
\end{equation} 
and $\langle \mathbf{S}_i \cdot \mathbf{S}_j \rangle_X =0$ 
for the other bonds.  
It is important to notice that these strong bonds 
are parts of the hexagon loop that connects the three 
tetrahedron units in $J'$ perturbation.  
The strong bonds form a Kekul\'e pattern of 
antiferromagnetic correlation on the hexagon loop. 
The same behavior was already discovered in the 
$S=\frac{1}{2}$ case, and 
this is the origin of stability of the state obtained 
~\cite{Tsunetsugu1,Tsunetsugu2,Moessner}.

\subsection{Solution in the bulk}
\label{subsec:solution}
Let us examine the stability of this configuration when 
the fourth unit $D$ is attached.  
This also gives a solution in the bulk.  
Since the effective interaction takes effect for each triad 
of tetrahedron units, this attachment increases the number 
of interacting triads from 1 to 4.  
For the solution obtained in Sec.~\ref{subsec:MFsol_triad}, 
I have found that all the new interactions relating to $D$ 
vanish as will be shown below.  
The mean field at the new unit $D$ is 
\begin{eqnarray}
&&\mathbf{h}_D = 
-
\mathbf{e}_0 
\bigl({\textstyle \frac{5}{4}} - \mathbf{e}_1 \!
       \cdot \! \langle \bm{\tau}_C \rangle\bigr)
\bigl({\textstyle \frac{5}{4}} - \mathbf{e}_2 \!
       \cdot \! \langle \bm{\tau}_B \rangle\bigr)
-
\mathbf{e}_1
\bigl({\textstyle \frac{5}{4}} - \mathbf{e}_2 \!
       \cdot \! \langle \bm{\tau}_A \rangle\bigr)
\bigl({\textstyle \frac{5}{4}} - \mathbf{e}_0 \!
       \cdot \! \langle \bm{\tau}_C \rangle\bigr)
\nonumber\\
&&\hspace{1.1cm} 
-
\mathbf{e}_2
\bigl({\textstyle \frac{5}{4}} - \mathbf{e}_0 \!
       \cdot \! \langle \bm{\tau}_B \rangle\bigr)
\bigl({\textstyle \frac{5}{4}} - \mathbf{e}_1 \!
       \cdot \! \langle \bm{\tau}_A \rangle\bigr) , 
\end{eqnarray}
and this vanish if $\langle \bm{\tau} \rangle_X $ ($X=A,B,C$) are fixed 
as before, because each 
$(\frac{5}{4} - \mathbf{e}_m \cdot \langle \bm{\tau}_X \rangle )$ is zero 
in the equation above.  
Therefore, $\langle \bm{\tau}_D \rangle$ is undetermined 
and may point to any direction with no energy cost.  
This does not affect the mean fields $\mathbf{h}_{A-C}$, 
as far as 
$\langle \bm{\tau}_{A-C} \rangle$ are fixed to the values 
of the three-unit solution.  
For example, in Eq.~(\ref{eq:hA}) for $\mathbf{h}_A$, 
the parts $\mathbf{e}_0 \cdot \langle \bm{\tau}_B \rangle$=
$\mathbf{e}_0 \cdot \langle \bm{\tau}_C \rangle$=%
$- \frac{1}{2} \langle \tau_{1,A} \rangle = \frac{5}{4}$ 
remove the contribution of $ \langle \bm {\tau}_D \rangle$. 
Therefore, this is a self-consistent solution 
of the four-unit problem, and its total energy of 
the four units is identical to that of the three-unit solution.
Thus the energy {\em per triad} increases.  
This situation happens in the $S=\frac{1}{2}$ case 
and its mean-field ground state in the four units $ABCD$ 
is continuously degenerate such that the direction of one 
$\langle \bm{\tau} \rangle$ is arbitrary ~\cite{Tsunetsugu1,Tsunetsugu2}.  

The situation completely changes in the $S$=$\frac32$ case. 
The solution above is one self-consistent solution, but 
there exists another solution with a lower energy. 
This difference comes from the 3-fold anisotropy in the 
$\bm{\tau}$ space in the $S=\frac{3}{2}$ case, 
{\em i.e.}, dependence of the ground state energy 
on the field direction $\epsilon_0 (\zeta )$.  
I have numerically solved the mean-field equations for the 4 
two-dimensional vectors $\{\langle \bm{\tau}_X \rangle \}$ ($X=A,B,C,D$),   
and found that the ground state is unique 
except 12-fold degeneracy due to the $T_d$ symmetry.  
All the 12 solutions have paired order parameters. 
One solution has the following ground state at each unit
\begin{subequations}
\begin{align}
\psi_0 (A) = \psi_0 (D) 
&= \sqrt{0.2902} \, \Phi_{A_1} + \sqrt{0.0064} \, \Phi_{A_2} 
\nonumber\\
&\phantom{=} + \sqrt{0.7034} \, 
\bigl( \cos \xi_{AD} \, \Phi_{E_u} + \sin \xi_{AD} \, \Phi_{E_v} \bigr), 
\ \ \ (\xi_{AD} = -0.3473 \pi )
\\
\psi_0 (B) = \psi_0 (C) 
&= 
- {\textstyle \sqrt{\frac{3}{10}}} \, \Phi_{A_1} 
+ {\textstyle \sqrt{\frac{7}{10}}} \, \Phi_{E_u} . 
\end{align}
\end{subequations}
Its order parameters pair up as 
\begin{equation}
\langle \bm{\tau}_{A} \rangle 
=\langle \bm{\tau}_{D} \rangle 
\approx 
(-1.8969,-2.9187) = 
\langle \bm{\tau} (1.253\pi ) \rangle 
, \ \ 
\langle \bm{\tau}_{B} \rangle 
=\langle \bm{\tau}_{C} \rangle 
= ({\textstyle \frac72},0) 
=\langle \bm{\tau} (0) \rangle  . 
\end{equation}
and this corresponds to spin correlations 
\begin{equation}
\begin{array}{cccc}
\mbox{units} & 
\ \langle \mathbf{S}_1 \cdot \mathbf{S}_2 \rangle  \ & 
\ \langle \mathbf{S}_1 \cdot \mathbf{S}_3 \rangle  \ & 
\ \langle \mathbf{S}_1 \cdot \mathbf{S}_4 \rangle  \
\\
\hline
\mbox{$A$ and $D$} \ & 
-3.1469 & 
-2.8293 & 
 2.2261
\\[2pt]
\mbox{$B$ and $C$} &  
{\textstyle \frac{9}{4}}   & 
-3 & 
-3
\end{array} 
\label{eq:MFcorrelations}
\end{equation}
With these values, all the 4 three-body couplings in Eq.~(\ref{eq:EMF}) 
for different triads have a negative value 
and thus lower the energy from the value for a tetrahedron triad, 
$E_{\mathrm{MF}} / (4 J_{\mathrm{eff}}) \approx -84.986 < -52.73$.
See also Table \ref{table}.  
This energy is the value for an isolated cluster of 4 tetrahedron units.  
I emphasize again that this is also a mean-field ground state 
in the bulk, and the bulk energy per cubic unit cell is 
four times of this value. 
Therefore the energy per spin is 
$E_{\mathrm{MF}} (\mbox{per spin}) / (4 J_{\mathrm{eff}} )
= -84.986 \times \frac{4}{16} = -21.25$.
This value does not contain the part of constant 
energy shift of orders $J'^2 /J$ and $J'^3 /J^2$. 

Let me explain the degeneracy of this mean-field 
ground state.  
In the solution above, two $\langle \bm{\tau} \rangle$'s point along 
one of the trigonal axes $\mathbf{e}_0$.  
The other two $\langle \bm{\tau} \rangle$'s 
are slightly tilted from another trigonal axis $\mathbf{e}_2$, 
but that tilt is small (about $\frac{1}{60}\pi$).  
This solution is degenerate in two ways. 
First, it is degenerate with respect to 
change $\tau_2 \rightarrow - \tau_2$.  
Secondly, the choice of two tetrahedron units 
are also arbitrary. 
One should note however that the trigonal axis 
to which their $\langle \bm{\tau} \rangle$'s 
point depends on which units are chosen.  
For example, in the case shown above, the units 
$B$ and $C$ have $\langle \bm{\tau} \rangle$
pointing to $\mathbf{e}_0$, which is related 
to 1-2 and 3-4 site pairs as shown in 
Eq.~(\ref{eq:twospin}). 
This corresponds to the lattice structure, in which 
the $B$ unit is positioned from the nearest $C$ unit along 
the direction of either 1-2 or 3-4 bond.  
Since there are 6 ways of choosing two from 
$ABCD$, the total degeneracy of the mean-field 
ground state is $2 \times 6 $=12, and 
these solutions are related to each other 
by symmetry operations of the $T_d$ point group.  

\begin{table}[t]
\caption{Spin correlations between neighboring tetrahedron units 
in one mean-field ground state for the $S=\frac{3}{2}$ case. 
Each triple product corresponds to one term in $E_{\mathrm{MF}}$, 
and the four products are all negative.  
$X_0$ in the triple product 
$\pi (X_0 )$=%
$\langle \mathbf{S}_1 \cdot \mathbf{S}_2 \rangle_{X}$%
$\langle \mathbf{S}_1 \cdot \mathbf{S}_3 \rangle_{X'}$%
$\langle \mathbf{S}_1 \cdot \mathbf{S}_4 \rangle_{X''}$
denotes the label of missing sublattice 
$X_0 = \{ A, B, C, D \} - \{ X, X', X'' \}$. 
\label{table}
}
\begin{center}
\begin{tabular}{ccccccc}
\hline
$X$ &  $X'$ & $X''$ & 
\ \ $\langle \mathbf{S}_1 \cdot \mathbf{S}_2 \rangle_{X} \! $ &
\ \ $\langle \mathbf{S}_1 \cdot \mathbf{S}_3 \rangle_{X'} \! $ &
\ \   $\langle \mathbf{S}_1 \cdot \mathbf{S}_4 \rangle_{X''} \! $ & 
triple product
\\
\hline
A & B & C & 
\ $-3.1469$           & $-3.0000$ & $-3.0000$           
& $\pi (D)$=$-28.322$ \\ 
B & A & D & 
\ $\phantom{-}2.2500$ & $-2.8293$ & $\phantom{-}2.2261$ 
& $\pi (C)$=$-14.171$ \\
C & D & A & 
\ $\phantom{-}2.2500$ & $-2.8293$ & $\phantom{-}2.2261$ 
& $\pi (B)$=$-14.171$ \\
D & C & B & 
\ $-3.1469$           & $-3.0000$ & $-3.0000$           
& $\pi (A)$=$-28.322$ \\
\hline
\end{tabular}
\end{center}
\label{table:correlation32}
\end{table}

\section{Spin correlation in the mean-field ground state in 
the $S=\frac{3}{2}$ case}
\label{subsec:Scorr}
I now use two-spin correlations 
$\langle \mathbf{S}_i \cdot \mathbf{S}_j \rangle$ 
and reexamine symmetry breaking 
in the mean-field ground state obtained in the previous section.  
There are two types of these correlations: 
one type is the correlations on short bonds 
inside tetrahedron unit 
and the other is those on long bonds between neighboring units. 
As will be shown below, spin correlations are finite only 
between nearest-neighbor sites within the perturbative 
approach used in the present work.  
However, they manifest spontaneous breaking of the point group symmetry 
in spin-singlet order.  

Spin correlations inside tetrahedron unit are 
already obtained during the calculation of the mean-field 
ground state, and their values are listed in Eq.~(\ref{eq:MFcorrelations}).  
Recall that it is sufficient to see 
$ f^{(1j)}=\langle \mathbf{S}_1 \cdot \mathbf{S}_j \rangle$ 
($j$=2, 3, and 4) because of the conjugate-pair equivalence. 
In the mean-field ground state for the $S=\frac{3}{2}$ case, 
two pairs have antiferromagnetic correlations 
and the other one is ferromagnetic in all the units. 
In two units among the four ($B$ and $C$ in Eq.~(\ref{eq:MFcorrelations})), 
the two antiferromagnetic correlations have the identical 
value $-3$, while they differ in the other two units.  

\subsection{Spin correlations between neighboring tetrahedron units 
for general $S$} 
\label{subsec:intercorr}

Correlations between different units require a more elaborate calculation.  
This is because original spin degrees of freedom 
$\{ \mathbf{S}_j (\mathbf{r}) \}$ 
are traced out and the effective Hamiltonian has 
only spin-pair operators $\{ \bm{\tau} (\mathbf{r}) \}$ in each unit.  
In the effective Hamiltonian approach, 
correlations of traced-out degrees of freedom are to be 
calculated from the hybridization of ground-state 
wave function with excited states, and this generally 
needs additional careful perturbative calculation.  
For example, 
for the large-$U$ limit of the half-filled Hubbard model, 
Bulaevskii \textit{et al.}
derived an expression of charge density and current 
in terms of spin operators \cite{hubbard}.   
In the present case, we can circumvent complicated calculation 
and obtain result quickly.  
The technique to use is Hellmann-Feynman theorem \cite{Hellmann}. 
Let us temporarily 
generalize the original Hamiltonian such that weak couplings  
on long bonds are all different depending their positions,   
$J' \rightarrow J_{ij}' (\mathbf{r},\mathbf{r}')$, and then 
its ground state energy is a function of these parameters, 
$E_{\mathrm{gs}} 
[\{ J'_{ij} (\mathbf{r}, \mathbf{r}' ) \} ]$.  
Its derivative with respect to one parameter is the 
corresponding spin correlation, and we finally set 
all the parameters to a uniform value $J'$ to come 
back to the original homogeneous Hamiltonian 
\begin{equation} 
\langle 
\mathbf{S}_{j_1 } (\mathbf{r}_1 ) \cdot 
\mathbf{S}_{j_2} (\mathbf{r}_2 ) \rangle  
= 
\left. 
\frac{\partial %
E_{\mathrm{gs}} \bigl[\{ J'_{ij} (\mathbf{r}, \mathbf{r}' ) \} \bigr] }
{\partial J'_{j_1 , j_2} (\mathbf{r}_1 , \mathbf{r}_2 )} 
\right|_{\mathrm{all} \  J'_{ij} (\mathbf{r}, \mathbf{r}' )=J'}  . 
\label{eq:Hellmann}
\end{equation} 
The approximation in the present work replaces the ground state energy 
$E_{\mathrm{gs}} [ \cdots ]$ 
by $\varDelta E_{0}^{(2)} [ \cdots ] +\varDelta E_{0}^{(3)} [ \cdots ] +
E_{\mathrm{MF}} [ \cdots ]$. 
Here, the last term is the mean-field ground state energy 
of $H_{\mathrm{eff}}$ and the first two terms are the energy shift 
in the second- and third-order perturbation
\begin{subequations}
\begin{align}
&\varDelta E_{0}^{(2)} 
\bigl[ \bigl\{ J'_{ij} (\mathbf{r}, \mathbf{r}' ) \bigr\} \bigr] 
=
- \frac{\, 3c_0^2 \,}{2J} \sum_{\langle \mathbf{r}, \mathbf{r}' \rangle} 
J'_{ij} (\mathbf{r}, \mathbf{r}' )^2 , 
\\
&\varDelta E_{0}^{(3)} 
\bigl[ \bigl\{ J'_{ij} (\mathbf{r}, \mathbf{r}' ) \bigr\} \bigr] 
= 
\phantom{-} \frac{\, 9c_0^3 \,}{2J^2} \!
\sum_{\langle \mathbf{r}_1 ,\mathbf{r}_2 ,\mathbf{r}_3 \rangle} 
J'_{i,j} (\mathbf{r}_1 ,\mathbf{r}_2 )
J'_{j,k} (\mathbf{r}_2 ,\mathbf{r}_3 ) 
J'_{k,i} (\mathbf{r}_3 ,\mathbf{r}_1 ) .  
\end{align}
\end{subequations}
Recall that the combination of neighboring units, 
$\langle \mathbf{r}, \mathbf{r}' \rangle$ or 
$\langle \mathbf{r}_1 ,\mathbf{r}_2 ,\mathbf{r}_3 \rangle$, 
automatically fixes the positions of connected sites, $ij$ or $ijk$.  
These energy shifts contribute a homogeneous part of spin correlations 
on long bonds 
\begin{equation}
C_{\mathrm{inter}} \equiv 
\left. 
\frac{\partial ( \varDelta E_{\mathrm{0}}^{(2)} + \varDelta E_{\mathrm{0}}^{(3)} ) 
\bigl[ \bigl\{ J'_{ij} (\mathbf{r}, \mathbf{r}' ) \bigr\} \bigr] }
{\partial J'_{j_1 , j_2} (\mathbf{r}_1 , \mathbf{r}_2 )} 
\right|_{\mathrm{all} \  J'_{ij} (\mathbf{r}, \mathbf{r}' )=J'}  
\!\! = - 3 c_0^2 \, \rho + 9 c_0^3 \, \rho^2 , 
\end{equation} 
where $\rho \equiv J'/J$ and 
the fact that each long bond is a part of two 
triangular loops is used for the part of $\varDelta E_{0}^{(3)}$.  

Non-uniform correlations come from the mean-field energy 
$E_{\mathrm{MF}}[ \cdots ]$.  
Since we need a result only for its first order 
derivative, one can use for 
$E_{\mathrm{MF}} [ \cdots ]$ the value given from Eq.~(\ref{eq:EMF})
by replacing $J_{\mathrm{eff}}$ with those local values calculated from 
$J_{ij}' (\mathbf{r},\mathbf{r}')$'s
\begin{align}
E_{\mathrm{MF}} [ \{ J'_{ij}(\mathbf{r}, \mathbf{r}' ) \} ] 
&=  
\frac{1}{6J^2} 
\sum_{\langle \mathbf{r}_1 ,\mathbf{r}_2 ,\mathbf{r}_3 \rangle} 
J'_{i_1,j_1} (\mathbf{r}_1 ,\mathbf{r}_2 )
J'_{j_2,k_1} (\mathbf{r}_2 ,\mathbf{r}_3 ) 
J'_{k_2,i_2} (\mathbf{r}_3 ,\mathbf{r}_1 ) 
\nonumber\\
&\hspace{1cm} \times 
\bigl( -c_0 + \mathbf{e}_{0} \cdot 
       \langle \bm{\tau}_{\mathbf{r}_1} \rangle \bigr)
\bigl( -c_0 + \mathbf{e}_{1} \cdot 
       \langle \bm{\tau}_{\mathbf{r}_2} \rangle \bigr)
\bigl( -c_0 + \mathbf{e}_{2} \cdot 
       \langle \bm{\tau}_{\mathbf{r}_3} \rangle \bigr). 
\label{eq:EMFnonlocal}
\end{align}
This replacement is exact up to the first order 
in each coupling in $\{ J_{ij}' (\mathbf{r},\mathbf{r}' ) \}$.  
The positions of connected sites, $i_1 , \cdots, i_2$, are 
also uniquely determined by the choice of three units 
$\langle \mathbf{r}_1 ,\mathbf{r}_2 ,\mathbf{r}_3 \rangle$.  
Note that each weak coupling $J'_{ij}(\mathbf{r}, \mathbf{r}' )$ 
appears just in two terms in the sum above.  
It is helpful to notice that 
one does not need to consider the contribution 
of order parameter deformation 
$\partial \langle \bm{\tau}_{\mathbf{r}} \rangle / 
\partial J'_{ij}(\mathbf{r}_1,\mathbf{r}_2)$ 
in Eq.~(\ref{eq:Hellmann}).  
This is because the deformation couples to 
$\delta E_{\mathrm{MF}} / \delta \langle \bm{\tau}_{\mathbf{r}} \rangle$, 
and this vanishes since the mean-field solution  
minimizes $E_{\mathrm{MF}}$.  

In the present case, all the bonds connecting the same 
pair of units $XX'$ have an identical value of spin 
correlations.  
For example, in Fig.~\ref{fig:network}(a), $A1$ unit 
is connected to surrounding four $D$ units,  
and spin correlations are all the same: 
$\langle \mathbf{S}_2 (A1) \cdot \mathbf{S}_1 (D1) \rangle$= 
$\langle \mathbf{S}_4 (A1) \cdot \mathbf{S}_3 (D2) \rangle$= 
$\langle \mathbf{S}_3 (A1) \cdot \mathbf{S}_4 (D4) \rangle$= 
$\langle \mathbf{S}_1 (A1) \cdot \mathbf{S}_2 (D5) \rangle$.   
This value is given by summing the contributions of constant energy 
shift and the mean field energy
\begin{subequations}
\begin{align}
\langle \mathbf{S}_i (A) \cdot \mathbf{S}_j (D) \rangle 
&= 
C_{\mathrm{inter}} 
+ 
{\textstyle \frac{1}{6}} \rho^2 
\left[ 
\bigl( -c_0 + \mathbf{e}_{0} \cdot 
       \langle \bm{\tau}_{C} \rangle \bigr)
\bigl( -c_0 + \mathbf{e}_{1} \cdot 
       \langle \bm{\tau}_{D} \rangle \bigr)
\bigl( -c_0 + \mathbf{e}_{2} \cdot 
       \langle \bm{\tau}_{A} \rangle \bigr) \right.  
\nonumber\\
&
\hspace{2.2cm}
\left.  + 
\bigl( -c_0 + \mathbf{e}_{0} \cdot 
       \langle \bm{\tau}_{B} \rangle \bigr)
\bigl( -c_0 + \mathbf{e}_{1} \cdot 
       \langle \bm{\tau}_{A} \rangle \bigr)
\bigl( -c_0 + \mathbf{e}_{2} \cdot 
       \langle \bm{\tau}_{D} \rangle \bigr) \right] 
\\
&=  
C_{\mathrm{inter}} 
+ {\textstyle \frac{1}{6}} \rho^2 
\bigl[ \pi (B) + \pi (C) \bigr] , 
\end{align}
\end{subequations}
where 
$\pi (X)$'s are triple products 
defined in Table \ref{table:correlation32}. 
The part of $E_{\mathrm{MF}}$ contributes 
two terms $\pi (B)$ and $\pi (C)$ corresponding to 
2 hexagon loops shown in Fig.~\ref{fig:loops}. 

For correlations of other pairs of sublattices, 
similar results are obtained and the combinations 
of the three units on the right-hand side are 
easily read from Eq.~(\ref{eq:EMF}) 
\begin{equation}
I_{X_1 X_2} \equiv \langle \mathbf{S}_i (X_1) \cdot \mathbf{S}_j (X_2) \rangle 
= 
C_{\mathrm{inter}} + 
{\textstyle \frac{1}{6}} \rho^2 
\bigl[ \pi (Y_1 ) + \pi ( Y_2 ) \bigr] , \ \ \ (X_1 \ne X_2 ) 
\end{equation}
where $Y$'s are determined by the complementary condition 
$\{ Y_1 , Y_2 \} = \{ A, B, C, D \} - \{ X_1 , X_2 \}$.  
For example, $I_{AC}$ is related to $\pi (B) + \pi (D)$.  
This leads to an important identity for the spin correlations between 
neighboring units, and this is a relation about correlations 
on different pairs of long bonds 
\begin{equation}
I_{AB} + I_{CD} = I_{AC} + I_{BD} = I_{AD} + I_{BC} 
= 2 C_{\mathrm{inter}} +  {\textstyle \frac{1}{6}} \rho^2 \sum_{X=A}^D \pi (X) 
=: 2\bar{I}_{\mathrm{inter}} . 
\label{eq:identityI}
\end{equation}
One should note that this identity holds generally for 
any mean-field solution.  
In the solution obtained in 
the previous section for 
the $S=\frac{3}{2}$ case, the order parameters pair up as 
$\langle \bm{\tau}_A \rangle$=$\langle \bm{\tau}_D \rangle$ and 
$\langle \bm{\tau}_B \rangle$=$\langle \bm{\tau}_C \rangle$, 
which leads to 
$\langle \pi (A) \rangle$=$\langle \pi (D) \rangle$ and 
$\langle \pi (B) \rangle$=$\langle \pi (C) \rangle$, 
but this degeneracy is not necessary for the above identity.  

Like the case of the effective Hamiltonian, the results 
for the spin correlations obtained up to this stage 
hold for general $S$.  

\begin{figure}[tb]
\begin{center}
\includegraphics[height=5cm]{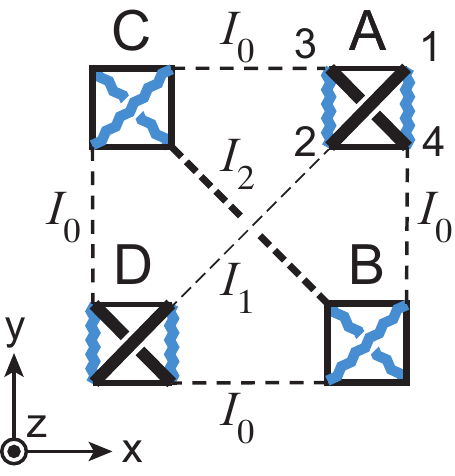}
\end{center}
\caption{Two-spin correlations in the $S=\frac32$ Heisenberg model 
on the breathing pyrochlore lattice. 
In each tetrahedron unit, 
antiferromagnetic correlations are shown by straight lines 
and their line width schematically shows 
$| \mathbf{S}_i \cdot \mathbf{S}_j |$, 
while zigzag lines show ferromagnetic correlations.  
Dashed lines show weak antiferromagnetic correlations 
between neighboring units, and they are of order $(J'/J)^1$ at most.  
\label{fig:order}}
\end{figure}

\subsection{Results of the $S=\frac{3}{2}$ case} 
\label{subsec:spin32}

I now apply the derived formula to the $S=\frac{3}{2}$ case 
and calculate spin correlations on long bonds connecting 
neighboring tetrahedron units.  
For this case, the constant value is 
$C_{\mathrm{inter}}=-\frac{75}{16} \rho + \frac{1125}{64} \rho^2 
\approx -4.6875 \rho + 17.578 \rho^2$.  
Triple products $\pi (X)$'s 
are already calculated and listed in Table~\ref{table}.  
Using these results up to $\rho^2$, the spin correlations 
on the long bonds are given as 
\begin{equation}
\left\{ 
\begin{array}{lll}
\displaystyle
I_{XY} = 
\langle \mathbf{S}_i (X) \cdot \mathbf{S}_j (Y) \rangle &
= -{\textstyle \frac{75}{16}} \rho + 10.4959 \rho^2 
  & =: I_0
, \hspace{0.5cm}
\bigl( X \in \{ A, D \}, \  
Y \in \{ B, C \} \bigr)
\\[5pt]
\displaystyle
I_{AD} = 
\langle \mathbf{S}_i (A) \cdot \mathbf{S}_j (D) \rangle & 
= -{\textstyle \frac{75}{16}} \rho + 12.8544 \rho^2 
  & =: I_{1} , 
\\[5pt]
\displaystyle
I_{BC} =
\langle \mathbf{S}_i (B) \cdot \mathbf{S}_j (C) \rangle &
= -{\textstyle \frac{75}{16}} \rho + \phantom{1} 8.1371 \rho^2 
  & =: I_{2}  . 
\end{array}
\right. 
\label{eq:Scorlong}
\end{equation}
This result shows that all the long bonds between 
tetrahedron units have an antiferromagnetic 
correlation, at least if the exchange coupling on long bonds 
is small enough $\rho \ll 1$.  
This is natural since the exchange coupling $J'$ on long bonds 
is antiferromagnetic.  
The $\rho^2$ terms contribute ferromagnetic correlations, 
and this comes from the constant energy shift $\varDelta E_{0}^{(3)}$.  
As $\pi (X) < 0$ for all $X$'s, 
$E_{\mathrm{MF}}$ yields antiferromagnetic contribution 
and this part is the origin of the difference between the three $I$'s.  
This result (\ref{eq:Scorlong}) is 
for one of the degenerate mean-field ground states, 
and spin correlations in other states are obtained by a symmetry operation 
in the $T_d$ point group.  

Figure~\ref{fig:order} illustrates spin correlations on short 
bonds in tetrahedron units as well as those on long bonds 
connecting neighboring units.  
The original tetrahedral symmetry $T_d$ 
of the units are lowered to 
$D_{2d}$ in the two units with higher symmetry 
and $D_2$ in the other two units. 
Note that $D_2$ is the lowest possible symmetry, 
since the equivalence relations should hold, 
and this is the point group symmetry of the entire 
system.

\subsection{Spin structure factor for the $S=\frac{3}{2}$ case}
\label{subsec:sq32}

Finally, I analyze spin structure factor. 
For that, I use the following definition  
\begin{equation}
S_{\mu \nu} (\mathbf{q})\equiv \frac{1}{N} 
\sum_{\mathbf{r},i}
\sum_{\mathbf{r}',j} 
\langle {S}_{i}^{\mu} (\mathbf{r}) {S}_{j}^{\nu} (\mathbf{r}') \rangle 
e^{-i \mathbf{q} \cdot 
(\mathbf{r} + \bm{\delta}_i - \mathbf{r}' - \bm{\delta}_j)} 
= {\textstyle \frac{1}{3}} \delta_{\mu \nu} S(\mathbf{q}) , 
\label{eq:defSq}
\end{equation} 
where $\bm{\delta}_i$ and $\bm{\delta}_j$ are the position 
of the $i$-th and $j$-th spin in tetrahedron unit.  
Instead of considering structure factor for each $ij$-pair, 
I alternatively define this for $\mathbf{q}$ in the extended Brillouin zone, 
not limited to the reduced zone.  
$S(\mathbf{q})$ has no Bragg peaks and is isotropic in the spin space, 
because of the absence of magnetic dipole order, 
and this is a smooth function of $\mathbf{q}$.  
Naturally, one divides $S(\mathbf{q})$ into two parts which 
correspond to correlations inside tetrahedron units and those between units: 
\begin{equation} 
S (\mathbf{q}) = S_{\mathrm{intra}} (\mathbf{q}) 
+ S_{\mathrm{inter}} (\mathbf{q}) . 
\end{equation}
The first part is calculated from spin correlations on short 
bonds, and the result is given by 
\begin{equation}
S_{\mathrm{intra}} (\mathbf{q}) = {\textstyle \frac{15}{4}} 
+ \bar{f}^{(12)} \gamma_{xy} (\mathbf{q} \, a_S ) 
+ \bar{f}^{(13)} \gamma_{zx} (\mathbf{q} \, a_S ) 
+ \bar{f}^{(14)} \gamma_{yz} (\mathbf{q} \, a_S ) , 
\label{eq:Sq}
\end{equation}
where the constant term is $S(S+1)$.  
Here, the form factor is 
$\gamma_{\mu \nu} (\bar{\mathbf{q}}) = 
\cos ( 2 \bar{q}_{\mu} )  \cos ( 2 \bar{q}_{\nu} )$
with $\sqrt2 a_s$ being the length of short bonds in tetrahedron units.  
$\bar{f}^{(ij)} = \frac14 \sum_{X=A}^{D} 
\langle \mathbf{S}_{i} \cdot \mathbf{S}_j \rangle_X$ 
is the average spin correlation between the spin pair $i$ and $j$ 
inside tetrahedron unit, 
and this is anisotropic in space depending on the bond direction; 
\begin{equation}
\bar{f}^{(12)}=-0.4485, \  
\bar{f}^{(13)}=-2.9147, \ 
\bar{f}^{(14)}=-0.3870, 
\end{equation}
and I have used the conjugate-pair equivalence, 
$\bar{f}_{12} = \bar{f}_{34}$ etc.  
To see the spatial anisotropy in more detail, I rewrite 
the $\mathbf{q}$-dependence and separate the part with cubic symmetry 
\begin{equation}
S_{\mathrm{intra}} (\mathbf{q}) = {\textstyle \frac{15}{4}} 
+ \bar{f}_{\mathrm{cub}} \, \gamma_{\mathrm{cub}} (\mathbf{q} \, a_S ) 
+ \bar{f}_u \gamma_{u} (\mathbf{q} \, a_S ) 
+ \bar{f}_v \gamma_{v} (\mathbf{q} \, a_S ) , 
\label{eq:Sq2}
\end{equation}
where 
\begin{equation} 
\gamma_{\mathrm{cub}} 
= \gamma_{xy} + \gamma_{yz} + \gamma_{zx} , \ 
\gamma_{u} 
= {\textstyle \frac{1}{\sqrt{6}}} 
( 2\gamma_{xy} - \gamma_{yz} - \gamma_{zx} ), \ 
\gamma_{v}
= {\textstyle \frac{1}{\sqrt{2}}} 
( \gamma_{yz} - \gamma_{zx} ), 
\end{equation} 
and the amplitudes are given by 
\begin{align}
&\bar{f}_{\mathrm{cub}} 
= {\textstyle \frac{1}{3}} 
(\bar{f}^{(12)} + \bar{f}^{(14)} + \bar{f}^{(13)} )
= - {\textstyle \frac{1}{3} } S (S+1) 
= - {\textstyle \frac{5}{4}} , 
\nonumber\\
&\bar{f}_{u} 
= {\textstyle \frac{1}{\sqrt{6}}} 
( 2\bar{f}^{(12)} - \bar{f}^{(14)} - \bar{f}^{(13)} )
= 0.9817, \hspace{0.7cm}
\bar{f}_{v}
= {\textstyle \frac{1}{\sqrt{2}}} 
( \bar{f}^{(14)} - \bar{f}^{(13)} )
= 1.7874, 
\end{align}
The cubic part of the $\mathbf{q}$-dependence is 
$\gamma_{\mathrm{cub}}$, and $\gamma_{u}$ and $\gamma_{v}$ 
represent spatially anisotropic components of 
spin correlations that transforms 
following $E$-irrep of the $T_d$ point group.  

The contribution of correlations between neighboring tetrahedron units 
is generally given as 
\begin{equation} 
S_{\mathrm{inter}} (\mathbf{q}) 
= {\textstyle \frac{1}{4}} 
\Bigl[ 
 ( I_{AB} + I_{CD} ) \gamma_{yz} ( \mathbf{q} \, a_L ) 
+( I_{AC} + I_{BD} ) \gamma_{zx} ( \mathbf{q} \, a_L ) 
+( I_{AD} + I_{BC} ) \gamma_{xy} ( \mathbf{q} \, a_L ) \Bigr] , 
\end{equation} 
where $\sqrt{2} a_L$ is the length of long bonds connecting 
tetrahedron units.  
$I_{XY}$'s values are listed in Eq.~(\ref{eq:Scorlong}).
The identity (\ref{eq:identityI}) guarantees that the three $\gamma$'s 
have a common amplitude, and its value is 
$\frac{1}{2} \bar{I}_{\mathrm{inter}}
= -  \frac{75}{32} \rho + 5.2480 \rho^2 $. 
Adding up the two parts, one obtains the final result of 
the spin structure factor 
\begin{align}
S(\mathbf{q} ) = {\textstyle \frac{15}{4}} 
&\Bigl[ 1 - {\textstyle \frac{1}{3}}  
            \gamma_{\mathrm{cub}} ( \mathbf{q} \, a_S ) 
 -  \bigl( {\textstyle \frac{5}{8}} \rho - 1.3995 \rho^2  \bigr) 
\gamma_{\mathrm{cub}} ( \mathbf{q} \, a_L )  
+ 0.2618 \gamma_{u} ( \mathbf{q} \, a_S )  
 + 0.4766 \gamma_{v} ( \mathbf{q} \, a_S )  \Bigr] , 
\end{align}
where the constant part $S(S+1)=\frac{15}{4}$ is factored out and 
again $\rho =\frac{J'}{J}$.

\section{Mean field ground state of the effective model in the $S=1$ case}
\label{sec:spin1}
Now that we have found that the mean-field ground state 
in the breathing pyrochlore model differ between the two cases 
of $S=\frac{1}{2}$ and $\frac{3}{2}$, a natural question 
arises about the intermediate case of $S=1$.  
It turns out that the result is similar 
to the $S=\frac{3}{2}$ case, and I will sketch calculations.  

In the case of $S=1$, the $S_{\mathrm{unit}}=0$ space at each 
tetrahedron unit has now dimension 3 and consists of 
basis states belonging to $A_1$- and $E$-irreps 
\begin{equation}
\Phi_{A_1} = {\textstyle \frac{1}{3}} 
\bigl( \sqrt5 \Phi_{0} + 2 \Phi_{2} \bigr) ,
\hspace{0.3cm} 
\Phi_{Eu} = {\textstyle \frac{1}{3}} 
\bigl( 2 \Phi_{0} - \sqrt5  \Phi_{2} \bigr) ,
\hspace{0.3cm} 
\Phi_{Ev} = - \Phi_{1} . 
\end{equation}
The effective Hamiltonian is the one in Eq.~(\ref{eq3:eff}) as 
before, but the constant is $c_0 = \frac{2}{3}$ and 
the two operators are now represented 
as follows 
with the local basis set $\{ \Phi_{A_1} , \Phi_{Eu},  \Phi_{Ev} \}$, 
\begin{equation}
\tau_1 = \frac{1}{3} \left[ \begin{array}{ccc} 
0           &   -2\sqrt{5} & 0 \\
-2\sqrt{5}  &    1         & 0 \\
0           &    0         & -1 
\end{array} \right], \ \ 
\tau_2 = \frac{1}{2} \left[ \begin{array}{ccc} 
0         &  0  & 2\sqrt{5}  \\
0         &  0  & 1 \\
2\sqrt{5} &  1  & 0 
\end{array} \right]. \ \ 
\end{equation}

\begin{figure}[t]
\begin{center}
\includegraphics[height=6cm]{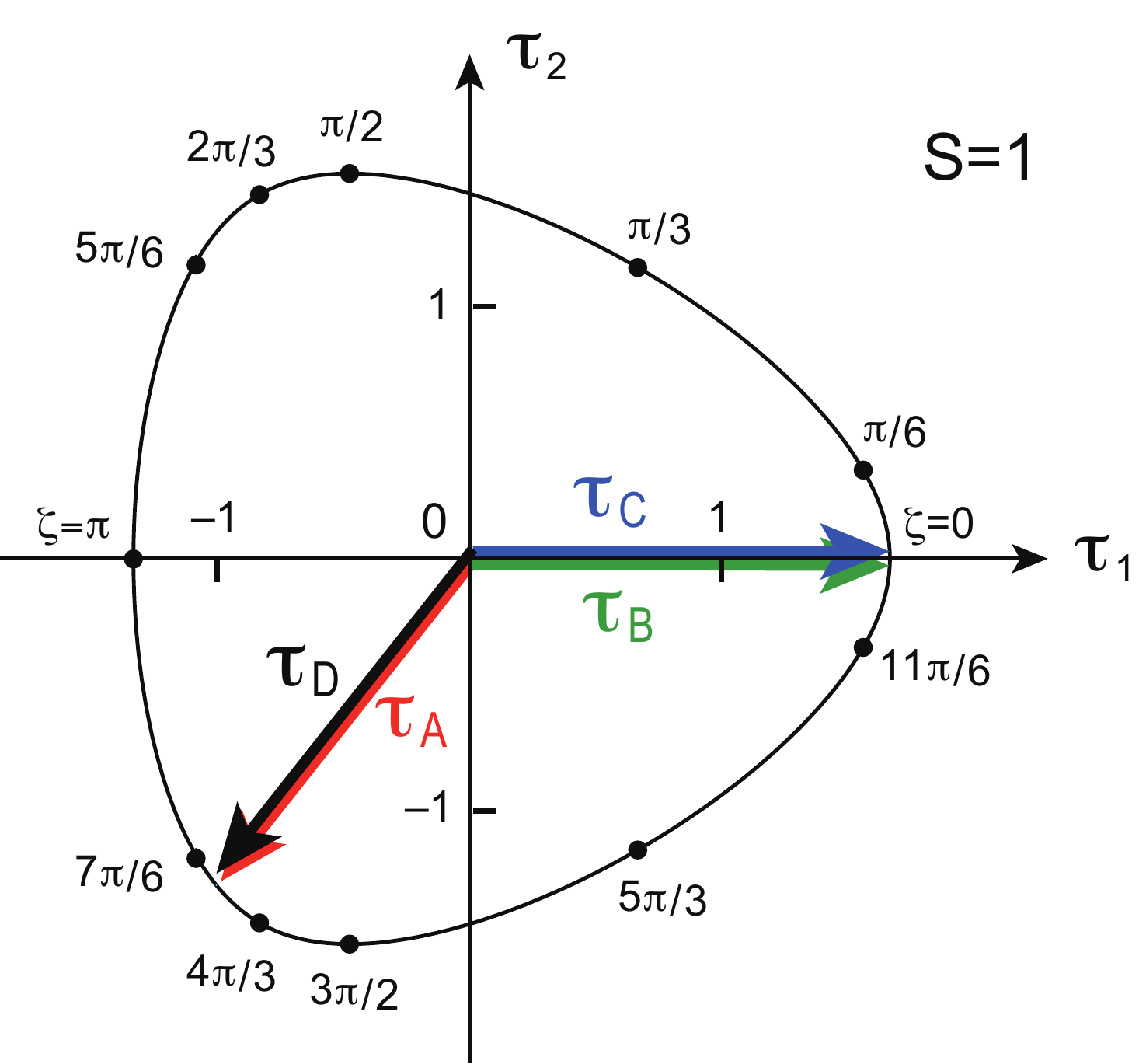}
\end{center}
\caption{
One mean-field ground state of the effective Hamiltonian for 
$S=1$.  
Trajectory is the ground state value of 
$\langle \bm{\tau}(\zeta ) \rangle$, and 
the expectation value $\langle \psi | \bm{\tau} | \psi \rangle$ 
calculated for any state $\psi$ 
should be located on or inside this trajectory.  
Two bounds are 
$\langle \tau_1 (\zeta =0) \rangle = \frac{5}{3}$ and 
$\langle \tau_1 (\pi ) \rangle = -\frac{4}{3}$.  
\label{fig:tauS1}}
\end{figure}

I have used the mean field approximation again for the $S=1$ 
case, and it goes exactly the same as before.  
A necessary calculation is a solution of the eigenvalue problem 
for a $3 \times 3$ matrix of 
the local mean field Hamiltonian (\ref{eq:mfHX}).  
I have diagonalized the reduced Hamiltonian $\bar{H}_{\mathrm{MF}} (\zeta )$, 
and found that its lowest eigenvalue is given as 
\begin{equation} 
\epsilon_0 (\zeta )= 
 - \frac{2 \sqrt{7}}{3} \cos 
\left[ \frac{1}{3} \cos^{-1} \left( 
\frac{10}{7^{3/2}} \cos 3 \zeta \right) \right] . 
\label{eq:e0S1}
\end{equation}
At the tetrahedron unit where the mean field $\mathbf{h}$ points to 
the direction $\mathbf{e}(\mathbf{\zeta })$, the order parameter 
is given by Eq.~(\ref{eq:OPzeta}) with this new result.  
Its trajectory upon varying $\zeta$ from 0 to $2\pi$ is 
plotted in Fig.~\ref{fig:tauS1}, and this has a shape 
of rounded triangle as in the $S=\frac{3}{2}$ case. 
This manifests anisotropy in the order parameter space, 
but the anisotropy is smaller compared to the $S=\frac{3}{2}$ case. 
As shown in Eq.~(\ref{eq:Ranis}), 
the anisotropy defined in the order parameter space is 
$R_{\mathrm{anis}}=\frac{5}{4}$, which is 
smaller than $R_{\mathrm{anis}} = \frac{7}{5}$ for $S=\frac{3}{2}$.  

The mean-field ground state for a tetrahedron triad 
does not depend on the $S$ value and the order parameters 
at the units $ABC$ have the same configuration as the 
one shown in Fig.~\ref{fig:mfsol}(a). 
A mean-field ground state in the tetrahedron quartet $ABCD$ 
is also a solution in the bulk.
Minimizing the mean field energy (\ref{eq:MFwf2}) with respect to 
the four order parameters $\langle \bm{\tau} \rangle$'s, 
I have searched ground states and found 12 solutions.  
One of them is 
\begin{equation}
\langle \bm{\tau}_{A} \rangle 
=\langle \bm{\tau}_{D} \rangle 
\approx 
(-1.0064,-1.2917) 
= 
\langle \bm{\tau} ( 1.215 \pi ) \rangle , 
\ \ 
\langle \bm{\tau}_{B} \rangle 
=\langle \bm{\tau}_{C} \rangle 
= ({\textstyle \frac53},0)
=\langle \bm{\tau} (0) \rangle .  
\end{equation}
These values are plotted in Fig.~\ref{fig:tauS1}, and 
this solution has the same nature 
of the solution in the $S=\frac{3}{2}$ case 
in Fig.~\ref{fig:mfsol}(b).  
The spin correlations are as follows
\begin{equation}
\begin{array}{cccc}
\mbox{units} & 
\langle \mathbf{S}_1 \cdot \mathbf{S}_2 \rangle  & 
\langle \mathbf{S}_1 \cdot \mathbf{S}_3 \rangle  & 
\langle \mathbf{S}_1 \cdot \mathbf{S}_4 \rangle  
\\
\hline
\mbox{$A$ and $D$} \ & 
-1.6731 & 
-1.2821 & 
 0.9552 
\\[2pt]
\mbox{$B$ and $C$} &  
 1                 & 
-{\textstyle \frac{3}{2}} & 
-{\textstyle \frac{3}{2}}  
\end{array} 
\end{equation}
Triple products in $E_{\mathrm{MF}}$ are all negative:  
$\pi (A) = \pi (D) = -3.7645$ and 
$\pi (B) = \pi (C) = -1.2247$.  
The relative difference between these two values 
is larger compared to the result for the $S=\frac{3}{2}$ case 
shown in Table ~\ref{table}.  
This is because $| \langle \bm{\tau}_{A,D} \rangle |$ is 
shorter than $|\langle \bm{\tau}_{B,C} \rangle |$ by about 
1.8\% here, while the reduction is only 0.5\% in the $S=\frac{3}{2}$ case.  

I have repeated the same calculation as in Sec.~\ref{subsec:sq32} 
and calculated spin structure factor for the $S=1$ case 
\begin{align}
S(\mathbf{q} ) = 2 
&\Bigl[ 1 
- {\textstyle \frac{1}{3} } \gamma_{\mathrm{cub}} ( \mathbf{q} \, a_S ) 
- \bigl( {\textstyle \frac{1}{3}} \rho - 0.4588 \rho^2  \bigr) 
  \gamma_{\mathrm{cub}} ( \mathbf{q} \, a_L )  
+ 0.2022 \gamma_{u} ( \mathbf{q} \, a_S )  
 + 0.3955 \gamma_{v} ( \mathbf{q} \, a_S )  \Bigr], 
\end{align}
where again $\rho =\frac{J'}{J}$.  
Compared to the $S=\frac{3}{2}$ case, 
the structure factor has smaller amplitudes in its 
anisotropic parts $\gamma_u$ and $\gamma_v$. 
This is related to the fact that 
the anisotropy $R_{\mathrm{anis}}$ in the order parameter 
is smaller for $S=1$.   

\section{Summary}
In this paper, I have studied the ground state 
of the antiferromagnetic spin-$S$ Heisenberg spin model 
on a breathing pyrochlore lattice, 
and examined a spontaneous breaking of lattice symmetry 
in the spin-singlet subspace.  
This lattice has two types of bonds, short and long, and 
the ratio of the corresponding exchange couplings $J'/J$ 
controls frustration.  
In the limit of $J'$=0, the ground state is thermodynamically degenerate 
and the main issue is what is the ground state when $0 < J' \ll J$ 
and what type of spatial pattern do spin correlations show 
in the symmetry broken ground state. 

Based on the third-order perturbation in $J'$, 
I have derived an effective Hamiltonian for general $S$, 
and examined spin-singlet orders with broken lattice symmetry. 
It is noticeable that the effective Hamiltonian 
has a form of three-tetrahedron interactions that is 
identical to the one of the $S$=$\frac{1}{2}$ case previously studied, 
and I have shown this with the help of newly found two identities 
of spin-pair operators.  
This Hamiltonian is represented in terms of two types 
of pseudospin operators $\bm{\tau}$, 
and they describe nonuniform correlations of four spins inside the unit.  
Despite of the identical form of the Hamiltonian, 
the dimension of $\bm{\tau}$'s local Hilbert space increases as 
$2S+1$ with spin. 
Since their matrix elements were not known, I have used 
algebras of four-spin composition and calculated these matrix 
elements for general $S$.  

Using these results, 
I have analyzed the effective model by a mean-field approximation 
and investigated its ground state for the special cases 
of $S$=$\frac{3}{2}$ and 1.  
I have found that the response of pseudospins has a $Z_3$ anisotropy 
in $(\tau_1,\tau_2 )$ space when $S > \frac{1}{2}$, and 
this has a critical effect on pseudospin orders.  
In contrast to the previous case of $S$=$\frac{1}{2}$, the ground state of 
four tetrahedron units has only a few stable configurations, 
and they have no further frustration 
when the configuration is repeated in space. 
Thus, this is the ground state of the entire system within 
the mean-field approximation.  
Actually, the ground state is uniquely determined in the sense that 
multiple solutions are related to each other by the lattice symmetry.  

I have studied in detail spatial pattern of spin correlations 
in the ground state.  
Each tetrahedron unit exhibits one of 
the two types of internal spin correlations as shown in Fig.~\ref{fig:order}. 
Among six bonds in each unit, four bonds have either strong or weak 
antiferromagnetic spin correlations, while the other two bonds have 
ferromagnetic correlations.  
As for correlations between different units, they 
are always antiferromagnetic.  
I have also calculated the spin structure factor $S (\mathbf{q})$, 
which may be compared to the energy-integrated value of 
neutron inelastic scattering.  
It is found that the amplitude of symmetry broken parts $\bar{f}_u$, 
$\bar{f}_v$ is comparable to the symmetric part $\bar{f}_{\mathrm{cub}}$. 

Quantum fluctuations between different units are neglected 
in the present work, but two features may justify this approximation.  
One is the three-dimensionality of the lattice, 
and quantum fluctuations are much smaller than in lower-dimensional cases.  
The second reason is the presence of the $Z_3$ anisotropy.   
A half of the local order parameters point to one of the 
directions that minimize the anisotropy energy, 
and the other half also only slightly tilt 
from another lowest-energy direction.  
Therefore, their fluctuations are expected not so large.  

Finally, I make a comment on the implication of the present result 
for interpreting experiment results in the Li$X$Cr$_4$O$_8$ compounds.  
It is not fair to compare the theoretical results of this work 
to experimental work, 
since the present theory does not take account of two important 
features in these materials. 
One is the magnetoelastic effects, i.e., coupling of spins and phonons.  
This is particularly important because the ground state of the spin 
system breaks the lattice translation and rotation symmetries. 
Coupling to the corresponding phonon modes should have a large 
contribution to the ground state energy, and this may change 
relative stability among different quantum states.  
The other is the spin anisotropy due to large $S$-value. 
It is legitimate to assume that orbital angular moments of Cr ions 
are quenched as a starting point, but fluctuations in Cr valency are 
not zero and this leads to anisotropic corrections to the Heisenberg 
couplings.  
I believe that despite these points the results and predictions 
in the present work are useful for discussing magnetic parts 
of possible symmetry breaking in the breathing pyrochlore materials.  


\section*{Acknowledgments}
The author is grateful to Yoshihiko Okamoto, Zenji Hiroi, You Tanaka, and 
Masashi Takigawa for enlightening discussions.

\appendix

\section{Classification of tetrahedron singlet states}
\label{sec:A1}

In this appendix, I classify the $(2S+1)$-fold 
singlet ground states in tetrahedron unit 
according to $T_d$ point group symmetry.  
This point group has one trivial and four other conjugacy 
classes of symmetry operations, 
and the nontrivial classes are represented by the following 
permutations of the four sites 
\begin{equation}
\mbox{IC$_4$: }
(1423) =P_2 , \ \ 
\mbox{C$_2$: }
(12)(34) =P_3, \ \ 
\mbox{$\sigma_d$: }
(12)(3)(4) =P_4, \ \ 
\mbox{C$_3$: }
(1)(234) =P_5, 
\label{eqA:permutation}
\end{equation}
where each parenthesis denotes the cyclic permutation of 
its contained sites \cite{Messiah}. 
For a permutation $P_n$ in each conjugacy class, 
its representation in terms of the singlet states is 
thus given by 
\begin{equation}
\bigl( P_n \bigr)_{l',l} = \langle \Phi_{l'} | P_n | \Phi_l \rangle .  
\end{equation}
Since operating permutation $P$ changes the way of spin combination 
in the basis states (\ref{eq2:CG4}), 
it is convenient to generalize the definition of basis states to 
\begin{equation}
\Phi_{l}^{(ij)(mn)} = 
\sum_{l_z=-l}^{l} 
C(l,l_z) 
\phi^{(ij)} (l,l_z ) \otimes 
\phi^{(mn)} (l,-l_z ) , 
\label{eqA:CG4}
\end{equation}
and the original ones are $\Phi_l = \Phi_l^{(12)(34)}$.  
The properties (\ref{eq:sym1}) and $C(l,-l_z)=C(l,l_z)$ 
lead to the following symmetry in the generalized basis states 
\begin{eqnarray}
&&\Phi_l^{(ij)(mn)} = (-1)^{2S+l} \Phi_l^{(ji)(mn)} 
= (-1)^{2S+l} \Phi_l^{(ij)(nm)} 
= \Phi_l^{(ji)(nm)} , 
\label{eqA:sym1}\\
&&\Phi_l^{(mn)(ij)} = \Phi_l^{(ij)(mn)} . 
\label{eqA:sym2}
\end{eqnarray}

It is ready to calculate the representation matrix 
for $P_3$, $P_4$, and $P_2$.  
Operating these permutations, we obtain 
\begin{equation}
P_3 \Phi_l = \Phi_l^{(21)(43)} = \Phi_l, \ \ 
P_4 \Phi_l = \Phi_l^{(21)(34)} = (-1)^{2S+l} \Phi_l, \ \ 
P_2 \Phi_l = \Phi_l^{(43)(12)} = (-1)^{2S+l} \Phi_l, 
\label{eqA:P}
\end{equation}
and therefore the representation matrix is 
$(P_2 )_{ll'}=(P_4 )_{ll'}=(-1)^{2S+l} \delta_{ll'}$ and 
$(P_3 )_{ll'}=\delta_{ll'}$.  
The character of their corresponding conjugacy classes 
is the trace of the representation matrices, and 
the result is 
$\chi (I C_4 ) = \chi (\sigma_d ) = 
\mbox{mod } (2S-1,2)$ 
and 
$\chi (C_2 )=2S+1$.  

As for $P_5$, we need to directly calculate the matrix element 
\begin{equation}
\bigl( P_5 \bigr)_{l' l}  
= \langle \Phi_{l'}^{(12)(34)} | \Phi_{l}^{(13)(42)} \rangle 
= (-1)^{2S+l} \langle \Phi_{l'}^{(12)(34)} | \Phi_{l}^{(13)(24)} \rangle .  
\end{equation}
This overlap of the two wavefunctions 
is a complicated factor that is related to the 
combination of four spins in two ways.  
Therefore it is possible to represent this with 
the Wigner's $9j$-symbol \cite{Messiah}
\begin{equation} 
\langle \Phi_{l'}^{(12)(34)} | \Phi_{l}^{(13)(24)} \rangle 
= (2l+1) (2l' +1) 
\left\{ \begin{array}{ccc} 
S & S & l' \\
S & S & l' \\
l & l & 0 \end{array} \right\}  . 
\label{eq:R9j}
\end{equation}
This can be further simplified by using a reduction 
formula of the $9j$-symbol to a $6j$-symbol \cite{Messiah}, 
and we obtain 
\begin{equation}
\bigl( P_5 \bigr)_{l' l} 
= \langle \Phi_{l'}^{(12)(34)} | P_5 | \Phi_{l}^{(12)(34)} \rangle 
=(-1)^{l'} \sqrt{(2l+1)(2l' +1)} 
\left\{ \begin{array}{ccc} 
S & S & l' \\
S & S & l  \end{array} \right\} . 
\label{eqA:6jP5}
\end{equation}
Finally, the character of the $C_3$ conjugacy class is 
\begin{equation}
\chi (C_3 ) = \mbox{Tr } P_5 
=\sum_{l=0}^{2S} (-1)^{l} (2l+1)
\left\{ \begin{array}{ccc} 
S & S & l \\
S & S & l  \end{array} \right\} 
= \mbox{mod } (2S-1,3)-1 , 
\end{equation}
where the last expression is derived based on 
the numerical results up to $S=6$.  

\begin{table}[tb]
\caption{Character table of the $T_d$ group.  
The last row is for the representation in terms of 
tetrahedron singlet states $\{ \Phi_l \}$.
See Ref.~\cite{group} for the irreps.} 
\label{tab:character}
\centering
\begin{tabular}{cccccc}
\hline
$T_d$  & $E$  & 6$IC_4$       & 
3$C_2$ & 6$\sigma_d$    & 8$C_3$ \\
\hline
A$_1$  & 1    & $\phantom{-}1$             & 
$\phantom{-}1$      & $\phantom{-}1$             & $\phantom{-}1$ \\
A$_2$  & 1    & $-1$          & 
$\phantom{-}1$      & $-1$          & $\phantom{-}1$ \\
E$_{\phantom{2}}$ 
       & 2    & $\phantom{-}0$             & 
$\phantom{-}2$      & 0             & $-1$ \\
T$_1$  & 3    & $\phantom{-}1$             & 
$-1$   & $-1$          & $\phantom{-}0$ \\
T$_2$  & 3    & $-1$          & 
$-1$   & 1             & $\phantom{-}0$ \\
\hline
$\Gamma_{\Phi}$ & $2S+1$ & $\mbox{mod } (2S-1,2)$ & 
$2S+1$   & $\mbox{mod } (2S-1,2)$ & $\mbox{mod } (2S-1,3)-1$ \\
\hline 
\end{tabular}
\end{table}

\begin{table}[t]
\caption{Multiplicity of the three irreps in $\Gamma_{\Phi}^{(S)}$ 
for $S \le 4$. 
$\chi_2 $=$\chi (IC_4 )$ and $\chi_5 $=$ \chi (C_3 )$. 
The multiplicity is 0 for the $T_1$- and $T_2$-irreps.   
}
\label{tab:multiplicity}
\centering
\begin{tabular}{rccccc}
\hline
$S$  & $\chi_2$  & $\chi_5$ & 
$n(\mathrm{A}_1 ; S)$ & $n(\mathrm{A}_2 ;S)$ & $n(\mathrm{E} ;S)$ \\
\hline
$\textstyle \frac{1}{2}$ & 0 & $-1$           & 0 & 0 & 1 \\[3pt]
1                        & 1 & $\phantom{-}0$ & 1 & 0 & 1 \\[3pt]
$\textstyle \frac{3}{2}$ & 0 & $\phantom{-}1$ & 1 & 1 & 1 \\[3pt]
2                        & 1 & $-1$           & 1 & 0 & 2 \\[3pt]
$\textstyle \frac{5}{2}$ & 0 & $\phantom{-}0$ & 1 & 1 & 2 \\[3pt]
3                        & 1 & $\phantom{-}1$ & 2 & 1 & 2 \\[3pt]
$\textstyle \frac{7}{2}$ & 0 & $-1$           & 1 & 1 & 3 \\[3pt]
4                        & 1 & $\phantom{-}0$ & 2 & 1 & 3 \\
\hline
\end{tabular}
\end{table}

The character of the conjugacy classes is listed in 
Table \ref{tab:character}, and it is interesting that 
the result is periodic in $S$ with period 3 except 
for the $E$ and $C_2$ classes.   

Using this character table, one can reduce the representation 
$\Gamma_{\Phi}^{(S)}$ by the tetrahedron singlet states $\{ \Phi_l \}$ to 
the irreps \cite{group}: 
$\Gamma_{\phi}^{(S)} = \sum_\Gamma n(\Gamma ; S ) \Gamma$ 
where $\Gamma$'s are the five irreps.   
The multiplicity is 
\begin{align}
&n (\mathrm{A}_{1} ; S) 
= \frac{2S+1 + 3\chi_2 + 2\chi_5 }{6}, \ \ 
 n (\mathrm{A}_{2} ; S) 
= \frac{2S+1 - 3\chi_2 + 2\chi_5 }{6}, \ \ 
\nonumber\\
&n (\mathrm{E} ; S) = \frac{2S+1- \chi_5 }{3}, \ \ 
 n (\mathrm{T}_{1} ; S) 
= n (\mathrm{T}_{2} ; S) = 0 , 
\end{align}
where $\chi_2 =\chi (IC_4 ) = \mbox{mod } (2S-1,2)$ and 
$\chi_5 = \chi (C_3 ) = \mbox{mod } (2S-1,3)-1$.  
The results are shown for $S \le 4$ in Table ~\ref{tab:multiplicity}. 
The periodicity in $\chi_{2}$ and $\chi_{5}$ 
leads to the following recursion formula 
\begin{equation}
n (\mathrm{A}_{1} ; S+3 ) = n (\mathrm{A}_{1} ; S ) +1 , \ \ 
n (\mathrm{A}_{2} ; S+3 ) = n (\mathrm{A}_{2} ; S ) +1 , \ \ 
n (\mathrm{E} ; S+3 ) = n (\mathrm{E} ; S ) +2  . 
\end{equation}

\section{Relation of $ F_{l' l}^{(12)} $, $ F_{l' l}^{(13)} $ and $ F_{l' l}^{(14)} $}
\label{sec:AF13}

In this appendix, I explain how to calculate 
matrix elements of $\mathbf{S}_1 \cdot \mathbf{S}_j $ 
between two basis states $\Phi_{l}$'s in the 
$S_{\mathrm{unit}}=0$ subspace in a tetrahedron unit. 
These states are defined in Eq.~(\ref{eq2:CG4}) 
in terms of two spin-pair wavefunctions 
$\phi^{(12)} (l,l_z)$ and $\phi^{(34)} (l,-l_z)$. 
Introducing composite spin operators 
$\mathbf{S}_{ij} \equiv \mathbf{S}_{i} + \mathbf{S}_{j}$, 
this $\phi^{(12)} (l,l_z)$ is an eigenstate of $\mathbf{S}_{12}^2$ 
with eigenvalue $l(l+1)$.
Summing up for $l_z$, the same is true about $\Phi_l$ and 
\begin{equation} 
l(l+1) \Phi_{l} = \mathbf{S}_{12}^2 \Phi_{l} 
= 2 \bigl[ S(S+1) + \mathbf{S}_{1} \cdot \mathbf{S}_{2} \bigr] \Phi_{l} 
= 2 \bigl[ S(S+1) + \mathbf{S}_{3} \cdot \mathbf{S}_{4} \bigr] \Phi_{l} . 
\label{eqA3:phi12}
\end{equation}
Therefore, $\Phi_l$ is an eigenstate of $\mathbf{S}_{1} \cdot \mathbf{S}_{2}$ 
and its eigenvalue is $\frac{1}{2}l(l+1)-S(S+1)$ and therefore 
\begin{equation} 
F_{l' l}^{(12)} \equiv 
\langle  \Phi_{l'} | {\textstyle \frac{1}{3}} S(S+1) 
+ \mathbf{S}_1 \cdot \mathbf{S}_2 | \Phi_{l} \rangle
= 
\bigl[ {\textstyle \frac{1}{2}} l(l+1) 
     - {\textstyle \frac{2}{3}} S(S+1) \bigr] \delta_{l' l} . 
\label{eqA3:S12}
\end{equation}
This matrix is traceless, $\sum_{l=0}^{2S} F_{ll}^{(12)}=0$.  

Matrix elements of $\mathbf{S}_1 \cdot \mathbf{S}_3 $ and 
$\mathbf{S}_1 \cdot \mathbf{S}_4 $ need 
more elaborate calculations, since the spin pair 1-3 or 1-4 
does not match the construction of the basis states $\Phi_l$'s. 
It is useful to notice that these are related to each other 
through the cyclic permutation $P_5$ introduced in Eq.~(\ref{eqA:permutation}). 
The operation $P_5$ changes the sites 2, 3, and 4 to 3, 4, and 2, respectively, 
while $P_5^{-1}=P_5^2$ changes to 4, 2, and 3.  
Therefore, one obtains 
\begin{equation} 
\mathbf{S}_1 \cdot \mathbf{S}_3 
= P_5 \bigl( \mathbf{S}_1 \cdot \mathbf{S}_2 \bigr) P_5^{-1}, \ \ 
\mathbf{S}_1 \cdot \mathbf{S}_4 
= P_5^{-1} \bigl( \mathbf{S}_1 \cdot \mathbf{S}_2 \bigr) P_5 .  
\label{eqA:SS234}
\end{equation}
This immediately leads to the relations between $F^{(1j)}$'s: 
\begin{equation} 
\mathsf{F}^{(13)} 
= P_5 \, \mathsf{F}^{(12)} \, {}^{t} \! P_5 , \ \ 
\mathsf{F}^{(14)} 
= {}^{t} \! P_5 \, \mathsf{F}^{(12)} \, P_5 ,  \ \ 
\label{eqA:F234}
\end{equation}
where the orthogonality of $P_5$ is used.  

Finally, I use these results for the special case of $S=\frac{3}{2}$. 
$\mathsf{F}^{(12)}$ is diagonal and obtained above. 
Nontrivial result is about $\mathsf{F}^{(13)}$.  
Evaluating the $6j$-symbols, the transformation matrix (\ref{eqA:6jP5}) 
is obtained with the basis set 
$\{ \Phi_0 , \cdots , \Phi_3 \}$ as 
\begin{equation}
P_5  = \frac{1}{20} \left[ 
\begin{array}{cccc} 
-5 &          5\sqrt{3}    & -5\sqrt{5} & 5\sqrt{7} \\
-5\sqrt{3} &  11           & -\sqrt{15} & - 3 \sqrt{21} \\
-5 \sqrt{5} & \sqrt{15}    &  15        & \sqrt{35} \\ 
-5 \sqrt{7} & -3 \sqrt{21} & -\sqrt{35} & - 1
\end{array} 
\right] . 
\end{equation}
Using this, the first relation in Eq.~(\ref{eqA:F234}) leads to 
\begin{equation} 
\mathsf{F}^{(13)} 
= 
\frac{1}{20} \left[ 
\begin{array}{cccc} 
25 & -25\sqrt{3} & 0 & 0 \\
-25\sqrt{3} & 15 & -8\sqrt{15} & 0 \\
0 & -8\sqrt{15} & -5 & -3\sqrt{35} \\ 
0 &  0 & -3\sqrt{35} & - 35 
\end{array} 
\right] . 
\end{equation}
Its diagonal part is $- \frac{1}{2} \tau_1$,  
while the off-diagonal part is $\frac{\sqrt{3}}{2} \tau_2$.   
This result of $\tau_2$ agrees with the direct calculation 
(\ref{eq:tau2S}).

\end{document}